\documentclass[10pt]{iopart}
\usepackage{graphicx}
\eqnobysec
\def\stackunder#1#2{\mathrel{\mathop{\;#1\;}\limits_{#2}}}
\newcommand{\beq}{\begin{equation}}
\newcommand{\beqa}{\begin{eqnarray}}
\newcommand{\eeq}{\end{equation}}
\newcommand{\eeqa}{\end{eqnarray}}

\newcommand{\abs}[1]{\vert#1\vert}
\renewcommand{\bar}[1]{{\overline{#1}}}
\newcommand{\cotanh}{\mathop{{\rm cotanh}}}
\newcommand{\dd}{{\rm d}}
\newcommand{\eps}{\varepsilon}
\newcommand{\g}{\gamma}
\newcommand{\half}{{\textstyle{\frac{1}{2}}}}
\newcommand{\ii}{{\rm i}}
\newcommand{\norm}[1]{\Vert#1\Vert}
\newcommand{\p}{\psi}
\newcommand{\var}{\mathop{{\rm var}}}
\newcommand{\Ai}{{\rm Ai}}
\newcommand{\Bi}{{\rm Bi}}
\renewcommand{\H}{{\cal H}}
\renewcommand{\P}{{\cal P}}
\newcommand{\T}{{\cal T}}
\newcommand{\PT}{{\cal PT}}
\renewcommand{\Re}{\mathop{{\rm Re}}}

\begin{document}

\title{An investigation of ${\cal PT}$-symmetry breaking in~tight-binding chains}

\author{Jean-Marc Luck}

\address{Universit\'e Paris-Saclay, CNRS, CEA, Institut de Physique Th\'eorique,
91191~Gif-sur-Yvette, France}

\begin{abstract}
We consider non-Hermitian ${\cal PT}$-symmetric tight-binding chains
where gain/loss optical potentials of equal magnitudes
$\pm{\rm i}\gamma$ are arbitrarily distributed over all sites.
The main focus is on the threshold $\gamma_c$ beyond which ${\cal PT}$-symmetry is broken.
This threshold generically falls off as a power of the chain length,
whose exponent depends on the configuration of optical potentials,
ranging between 1 (for balanced periodic chains)
and 2 (for unbalanced periodic chains,
where each half of the chain experiences a non-zero mean potential).
For random sequences of optical potentials with zero average and finite variance,
the threshold is itself a random variable, whose mean value decays with exponent 3/2
and whose fluctuations have a universal distribution.
The chains yielding the most robust ${\cal PT}$-symmetric phase,
i.e., the highest threshold at fixed chain length,
are obtained by exact enumeration up to 48 sites.
This optimal threshold exhibits an irregular dependence on the chain length,
presumably decaying asymptotically with exponent 1, up to logarithmic corrections.
\end{abstract}

\ead{\mailto{jean-marc.luck@ipht.fr}}

\maketitle

\section{Introduction}
\label{intro}

Non-Hermitian Hamiltonians involving complex optical potentials
have been used in various guises for a long time,
in order to model either inelastic scattering or absorption,
in analogy with the use of complex refraction indices in optics.
Non-Hermitian operators have gained much attention lately,
both in classical and quantum physics
(see~\cite{rev} for a recent comprehensive review).
Non-Hermitian Hamiltonians generically have complex spectra,
with a resulting non-unitary dynamics.
It was however emphasized by Bender and collaborators~\cite{ben1,ben2},
generalizing an observation made by Bessis and Zinn-Justin
on the Lee-Yang Hamiltonian $\H=p^2+\ii x^3$ (see~\cite{jzjj}),
that a whole class of non-Hermitian Hamiltonians may have real spectra.
These are the $\PT$-symmetric Hamiltonians,
which are invariant under the combined action
of spatial parity ($\P$) and time reversal~($\T$)
(see~\cite{benrev} for a full technical exposition,
and~\cite{bennews,natphys} for more accessible overviews).
The generic scenario is the following.
Let~$\H$ be a non-Hermitian $\PT$-symmetric Hamiltonian, obeying $[\H,\PT]=0$.
If $\H$ is close to being Hermitian,
in the sense that $\norm{\H^\dag-\H}$ is small enough,
$\H$ and $\PT$ have a complete set of common eigenvectors,
all energy eigenvalues of $\H$ are real,
and the dynamics is unitary.
The system is in a $\PT$-symmetric phase, and $\PT$-symmetry is said to be unbroken.
If $\norm{\H^\dag-\H}$ gets larger,
it may occur that some of the eigenvectors of $\H$ are not eigenvectors of $\PT$ any more.
This is possible since the~$\PT$ operator is not linear, but anti-linear
(see~\cite{benrev} for details).
The corresponding energy eigenvalues are complex,
so that the norm of a generic state vector blows up exponentially fast in time.
$\PT$-symmetry is said to be broken.
The number of pairs of complex conjugate energy eigenvalues usually increases
with $\norm{\H^\dag-\H}$.
The threshold for $\PT$-symmetry breaking corresponds to the very first occurrence
of such a pair of complex eigenvalues.

Among the vast variety of $\PT$-symmetric classical and quantum systems
that have been investigated so far in more than 3,000 published papers,
tight-binding models are definitely among the easiest ones to grasp,
at least on the theoretical side.
Such models describe e.g.~optical arrays consisting of coupled units,
some of them having gains and losses.
Most studies concern the one-dimensional geometry of
tight-binding chains~\cite{ref0,ref1,ref2,ref3,ref4,refflat,ref5,refort}.
Various cases of finite chains consisting of $N=2M$ sites, either pristine or disordered,
hosting one or several $\PT$-symmetric pairs of non-Hermitian impurities have been considered.
Scattering states and transport properties of $\PT$-symmetric chains
coupled to an environment through leads have also been considered~\cite{refA,refB,refC,refD}.
Most of these studies have dealt with the situation where all gains and losses
have the same magnitude~$\g$, i.e.,
the optical potentials of all sites with gain (respectively, with loss)
read~$+\ii\g$ (respectively,~$-\ii\g$).
The threshold for $\PT$-symmetry breaking is then a single number~$\g_c$,
whose dependence on various model parameters has been explored in some detail
in various circumstances.
The presence of a single $\PT$-symmetric pair of sites with gain and loss
already yields a rich phenomenology~\cite{ref1,ref2,ref3,refort}.
The situation where each site bears a random gain/loss optical potential $\pm\ii\g$
has also been considered~\cite{ref5}.
The dependence of the threshold on the chain length
has been found to strongly differ from one situation to another:
it reaches a finite limit in the case of a single pair of $\PT$-symmetric impurities
at the endpoints of a long pristine chain~\cite{ref1,ref3},
whereas it is exponentially small in~$M$ in the insulating regime induced by Anderson localization,
i.e., when the on-site energies have strong enough random real parts~\cite{ref0,ref4,ref5}.
For $\PT$-symmetric gain/loss disorder alone,
numerical simulations have revealed
that the threshold has a smooth distribution
whose mean value falls off slowly as a function of the chain~length~\cite{ref5}.

The aim of this paper is to put some of the findings recalled just above in a broader perspective,
by providing a systematic study of the threshold
of $\PT$-symmetric tight-binding chains of length $N=2M$
where gain/loss optical potentials are distributed over all sites
in a periodic, random, or any other fashion.
For definiteness, we focus our attention onto a minimal setting,
where all gains and losses have the same magnitude~$\g$,
whereas the chains are otherwise pristine,
in the sense that there are no other sources of inhomogeneity,
besides the optical potentials
$\ii\g\eps_n$, with $\eps_n=\pm1$.
This model easily lends itself to an exact enumeration of all possible sequences~$\eps_n$
of reduced optical potentials for a given chain size.
In this setting, the threshold for $\PT$-symmetry breaking is a single number~$\g_c$,
depending only on the arrangement of the reduced optical potentials~$\eps_n$.
Right at this threshold, the spectrum 
usually exhibits two symmetry-related exceptional points,
i.e., two (usually twofold) degenerate eigenvalues at opposite energies
($\pm E_c$, with $E_c>0$),
or, albeit more rarely, a single exceptional point,
i.e., a single (again usually twofold) degenerate eigenvalue at zero energy.
More complex patterns of exceptional points may however occur at $\g_c$.
We shall meet two examples of such situations in the course of this work.
In Section~\ref{robust} we provide an example with $M=6$,
where four twofold degeneracies take place simultaneously at non-zero energies given by~(\ref{e1e2}).
In \ref{appg1g2} it is shown that for $M=4$ with generic optical potentials
there are four isolated points in the $\g_1$--$\g_2$ plane
for which a fourfold degeneracy occurs at threshold at zero energy.

Throughout this work,
the main emphasis is on the scaling behavior of the threshold~$\g_c$ as a function of the chain length
for various classes of configurations of optical potentials.
Section~\ref{theory} contains the theoretical framework of the model,
including the transfer-matrix approach and the discrete Riccati formalism.
General results are presented in Section~\ref{general},
including explicit results for all chains up to $M=3$ (Section~\ref{short})
and an overview of the behavior of the threshold for various classes of sequences
(Section~\ref{pano}).
More specific results for these classes of sequences are derived in the subsequent sections.
This detailed study includes diblock chains
and other unbalanced periodic chains (Section~\ref{diper}),
alternating chains and other balanced periodic chains (Section~\ref{alper}),
random chains (Section~\ref{ran}),
and the chains yielding the most robust $\PT$-symmetric phases,
i.e., the largest threshold at fixed chain length (Section~\ref{robust}).
Section~\ref{disc} is devoted to a brief discussion.
Three appendices contain more technical matters.

\section{Theoretical framework}
\label{theory}

\subsection{The model}
\label{model}

The focus of this work is on the time-independent
$\PT$-symmetric tight-binding equation
\beq
\p_{n-1}+\p_{n+1}+\ii\g\eps_n\p_n=E\p_n
\label{tbe}
\eeq
on a finite open chain of $N=2M$ sites,
with Dirichlet boundary conditions
\beq
\p_0=\p_{2M+1}=0.
\label{dir}
\eeq
Each site bears a purely imaginary gain/loss optical potential of magnitude $\g$,
whereas $\eps_n=\pm1$ is an arbitrary sequence of signs,
with $\eps_n=+1$ (abbreviated to $\eps_n=+$) corresponding to gain,
and $\eps_n=-1$ (abbreviated to $\eps_n=-$) to loss.
We always assume that the model is $\PT$-symmetric.
This translates to the constraint
\beq
\eps_{2M+1-n}=-\eps_n,
\label{pt}
\eeq
imposing that the sequence of signs is of the form
\beq
\eps_1,\dots,\eps_{2M}=\underbrace{\eps_1,\eps_2,\dots,\eps_M},
\underbrace{-\eps_M,\dots,-\eps_2,-\eps_1}.
\label{palin}
\eeq
In the following,
$(\eps_1,\dots,\eps_{2M})$ will be referred to as the sequence defining the chain,
and $(\eps_1,\dots,\eps_M)$ as the corresponding half-sequence.

For a given $\PT$-symmetric sequence,
the spectrum of~(\ref{tbe}) consists of $2M$ energy eigenvalues $E_a$,
which can either be real or occur in complex conjugate pairs
(see Section~\ref{ric} for more details).
This spectrum is invariant under $\P$ and $\T$ separately,
so that one can choose $\g\ge0$ without loss of generality.
We are mainly interested in the threshold $\g_c$ associated with the occurrence
of the first pair of complex conjugate eigenvalues,
and especially in the dependence of $\g_c$ on the length $M$ and on the nature
(periodic, random, etc.) of the half-sequence.
It is worth recalling here that only $\PT$-symmetric sequences give rise to a non-zero $\g_c$.
This can be shown by means of first-order perturbation theory
(see~\cite{ref5}, and~\ref{apppert} for a detailed proof).

\subsection{Transfer-matrix approach}
\label{tm}

The transfer-matrix approach is a very useful tool
in the theory of one-dimensional disordered systems~\cite{CPV,alea,pendry,CT1,CT2}.
In the present situation, the tight-binding equation~(\ref{tbe}) can be recast as
\beq
\pmatrix{\p_{n+1}\cr\p_n}=T_n\pmatrix{\p_{n}\cr\p_{n-1}},\qquad
\pmatrix{\p_{n-1}\cr\p_n}=T_n\pmatrix{\p_{n}\cr\p_{n+1}},
\label{recm}
\eeq
where the transfer matrix $T_n$ associated with site $n$ reads
\beq
T_n=\pmatrix{E-\ii\g\eps_n&-1\cr1&0}.
\eeq
Iterating the recursion~(\ref{recm}) along the half-sequence
and along its $\PT$-symmetric partner,
using the Dirichlet boundary conditions~(\ref{dir})
and the $\PT$-symmetry constraint~(\ref{pt}),
we respectively obtain\footnote{Throughout this paper a star denotes complex conjugation.}
\beq
\pmatrix{\p_{M+1}\cr\p_M}=U_M\pmatrix{\p_1\cr0},\qquad
\pmatrix{\p_{M}\cr\p_{M+1}}=U^\star_M\pmatrix{\p_{2M}\cr0},
\eeq
with
\beqa
U_M=T_M\dots T_1=\pmatrix{a_M&b_M\cr c_M&d_M},
\nonumber\\
U^\star_M=T_{M+1}\dots T_{2M}=\pmatrix{a^\star_M&b^\star_M\cr c^\star_M&d^\star_M}.
\eeqa
We have therefore
\beq
\p_M=c_M\p_1=a^\star_M\p_{2M},\qquad
\p_{M+1}=a_M\p_1=c^\star_M\p_{2M}.
\eeq
The quantization condition which determines the spectrum of~(\ref{tbe}) therefore reads
\beq
a_Ma^\star_M=c_Mc^\star_M,\qquad\hbox{i.e.,}\quad\abs{a_M}=\abs{c_M}.
\label{qac}
\eeq
We have $c_n=a_{n-1}$, whereas the entries $a_n$ obey the linear recursion
\beq
a_n=(E-\ii\g\eps_n)a_{n-1}-a_{n-2},
\eeq
with $a_0=1$ and $a_{-1}=0$.

\subsection{Discrete Riccati formalism}
\label{ric}

Discrete Riccati variables are another classic tool
of the theory of one-dimensional disordered systems~\cite{CPV,alea,pendry,CT1,CT2}.
In the present situation, they read
\beq
R_n=\frac{a_n}{a_{n-1}}.
\eeq
These ratios obey the non-linear recursion
\beq
R_n=E-\ii\g\eps_n-\frac{1}{R_{n-1}},
\label{recr}
\eeq
known as a discrete Riccati equation, with initial value $R_0=\infty$.
In terms of these variables, the quantization condition~(\ref{qac}) reads
\beq
R_MR^\star_M=1,\qquad\hbox{i.e.,}\quad\abs{R_M}=1.
\label{qr}
\eeq

The above condition yields after reduction a polynomial equation
of the form $P(E,\g)=0$,
where the characteristic polynomial $P$ has degree $2M$ in $E$ and in $\g$,
has real coefficients, and is even in each of the variables $E$ and $\g$
(see examples in Section~\ref{short}).
These properties have several consequences.

For a generic value of $\g$,
the spectrum of~(\ref{tbe}) consists of $2M$ distinct energy eigenvalues,
which are the zeros of $P$.
If $E$ is an eigenvalue, so are $E^\star$, $-E$ and $-E^\star$.
These four numbers degenerate into two either if $E$ is real (and so $E^\star=E$)
or if $E$ is imaginary (and so $E^\star=-E$).
They degenerate into one if $E=0$.

At the threshold $\g_c$,
there are usually two (usually twofold) degenerate eigenvalues at opposite real energies
($\pm E_c$, with $E_c>0$),
or, albeit more rarely, a single (usually twofold) degenerate eigenvalue at zero energy.
As announced in the Introduction,
more complex degeneracy patterns may take place at $\g_c$.

The recursion~(\ref{recr}) implies
\beq
\abs{R_n}+\frac{1}{\abs{R_{n-1}}}\ge\sqrt{E^2+\g^2}
\eeq
whenever $E$ is real.
If the r.h.s.~of the above inequality is larger than 2,
it can be shown by recursion that $\abs{R_n}>1$ for all $n$,
so that the quantization condition~(\ref{qr}) cannot be satisfied.
In other words, all real eigenvalues of~(\ref{tbe}) obey
\beq
\abs{E}\le\sqrt{4-\g^2},
\eeq
independently of the length $M$.
The real spectrum is therefore reduced
with respect to that of a free tight-binding particle ($\abs{E}\le2$).
It shrinks down to zero as $\g\to2$.
In particular, all sequences have $\g_c<2$.

\section{General results}
\label{general}

\subsection{Shortest chains}
\label{short}

As already said,
the spectrum of~(\ref{tbe}) is invariant under $\P$ and $\T$ separately.
As a consequence, besides choosing $\g\ge0$,
one can fix $\eps_1=+$, and so $\eps_{2M}=-$, without loss of generality.
There are therefore $2^{M-1}$ essentially different $\PT$-symmetric sign sequences
on a chain of length $N=2M$, each of them being encoded in a half-sequence of length $M$
starting with $+$.

\begin{figure}[!ht]
\begin{center}
\includegraphics[angle=0,width=.4\linewidth,clip=true]{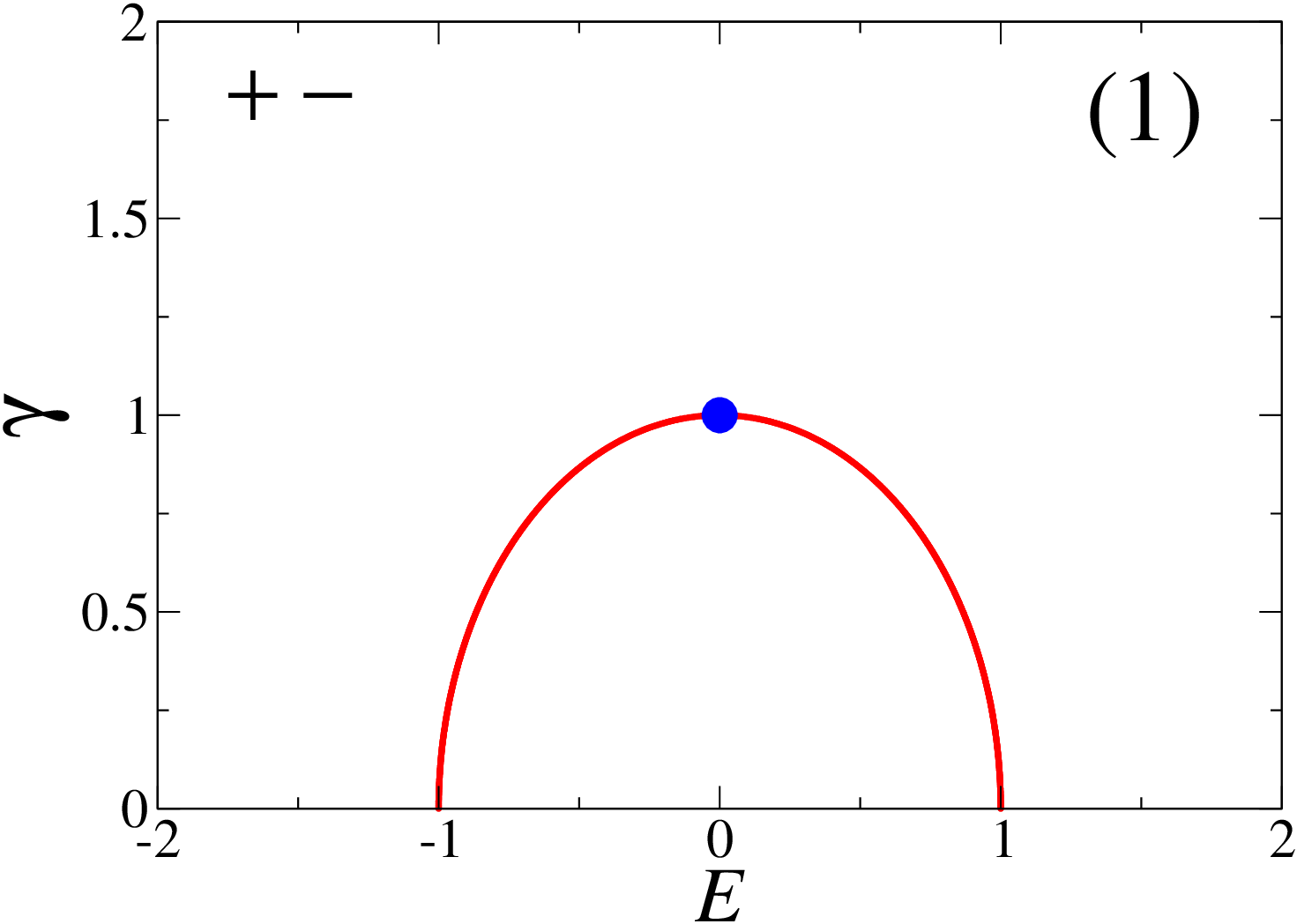}
\vskip 4pt
\includegraphics[angle=0,width=.4\linewidth,clip=true]{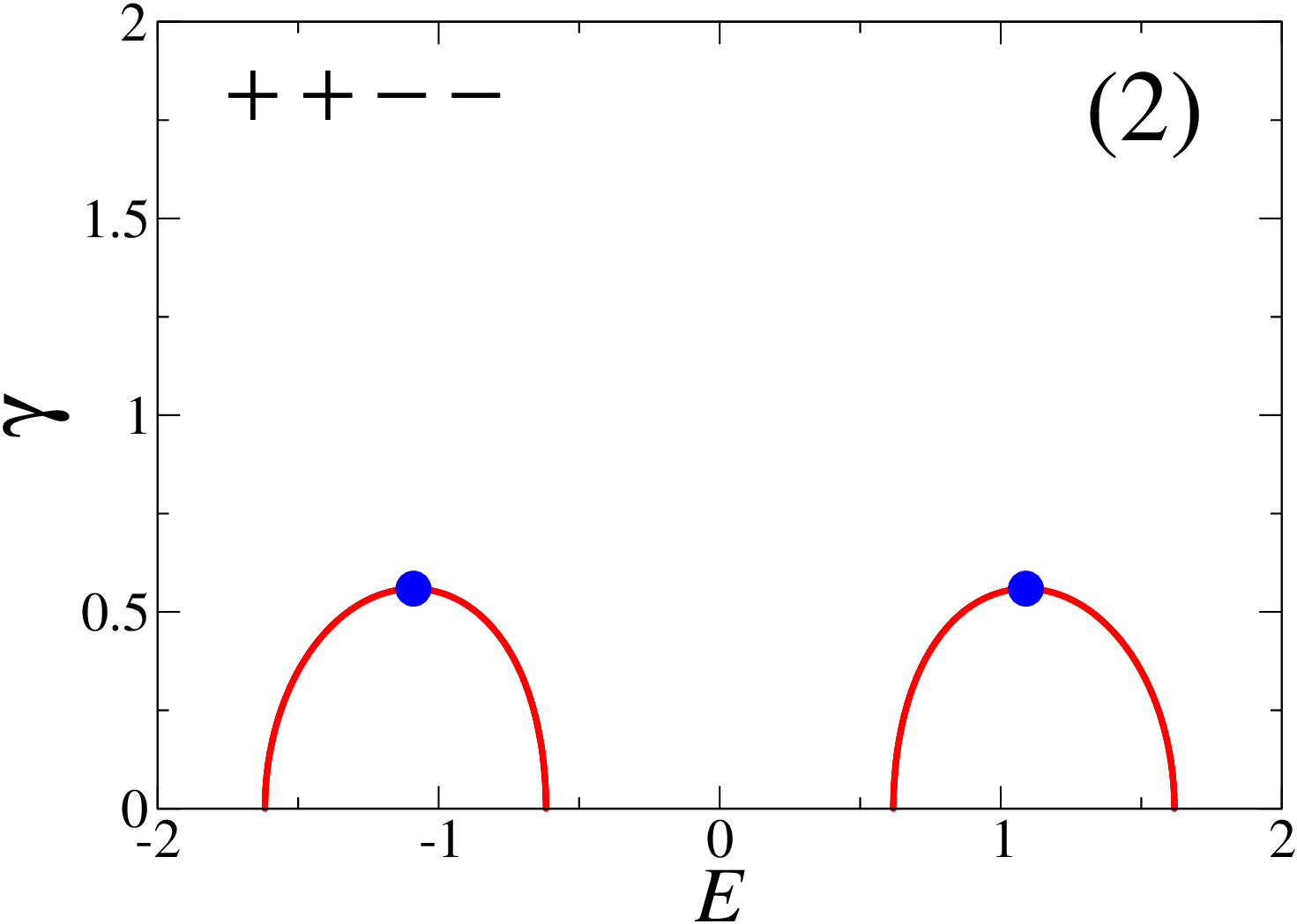}
\hskip 8pt
\includegraphics[angle=0,width=.4\linewidth,clip=true]{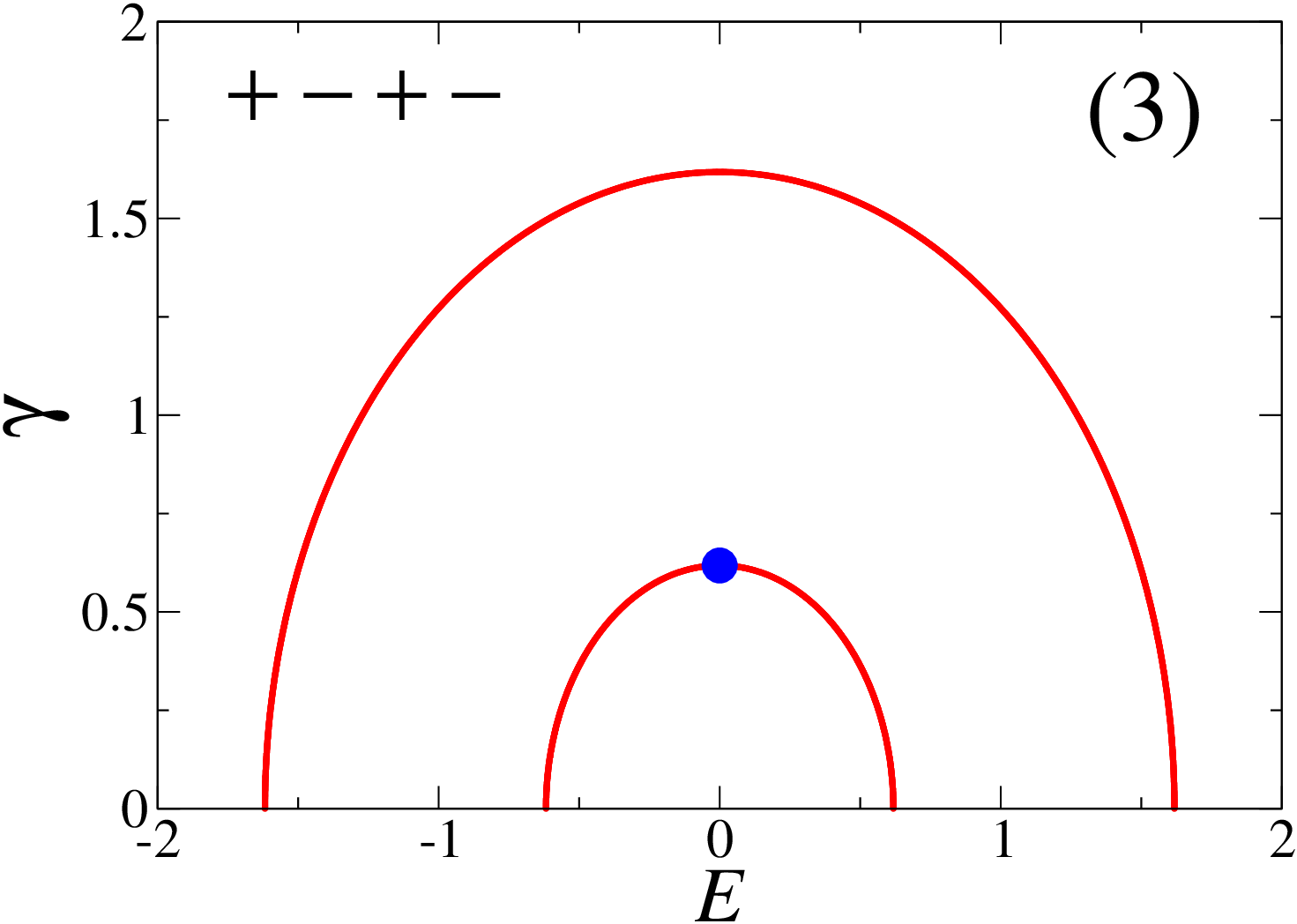}
\vskip 4pt
\includegraphics[angle=0,width=.4\linewidth,clip=true]{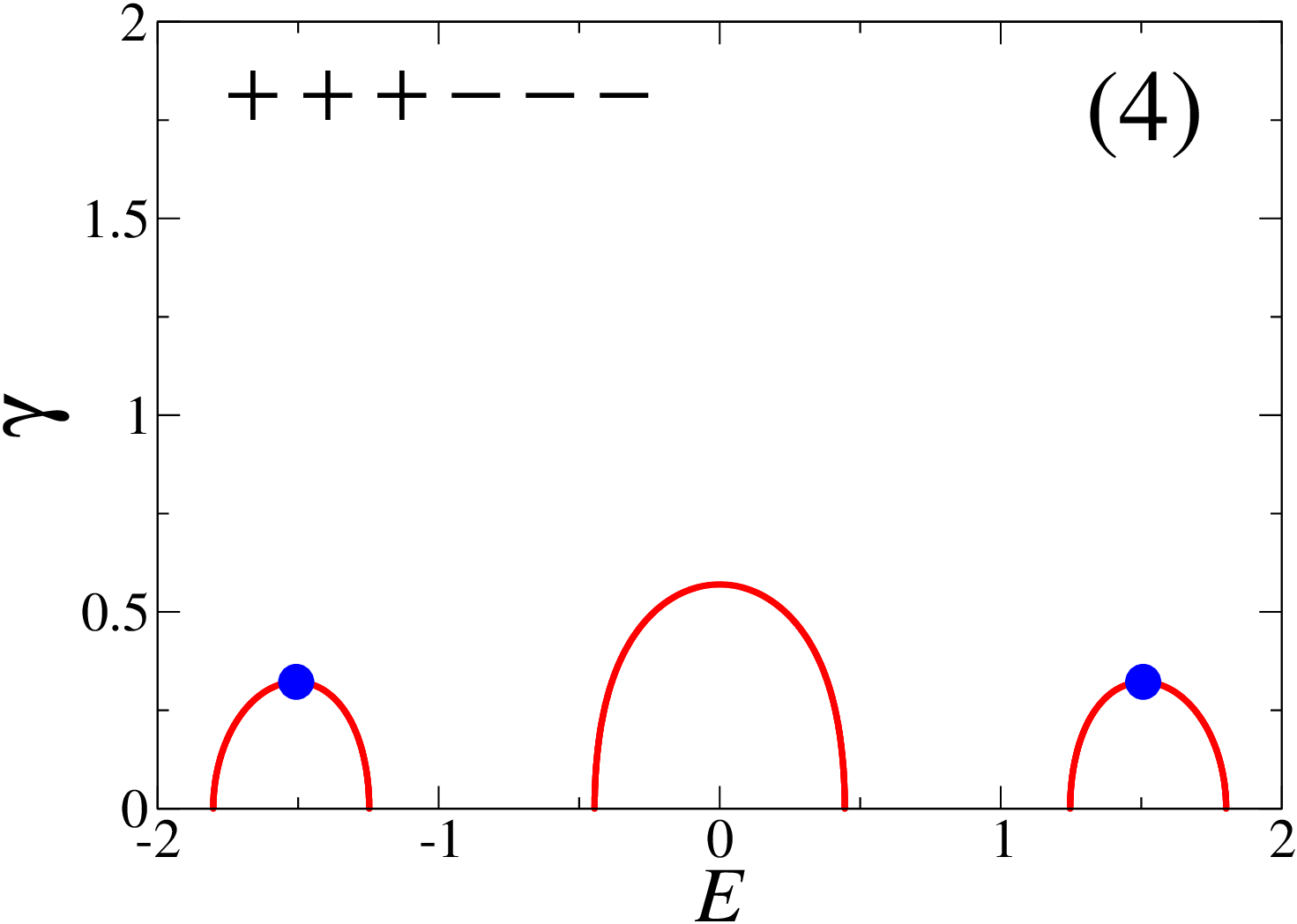}
\hskip 8pt
\includegraphics[angle=0,width=.4\linewidth,clip=true]{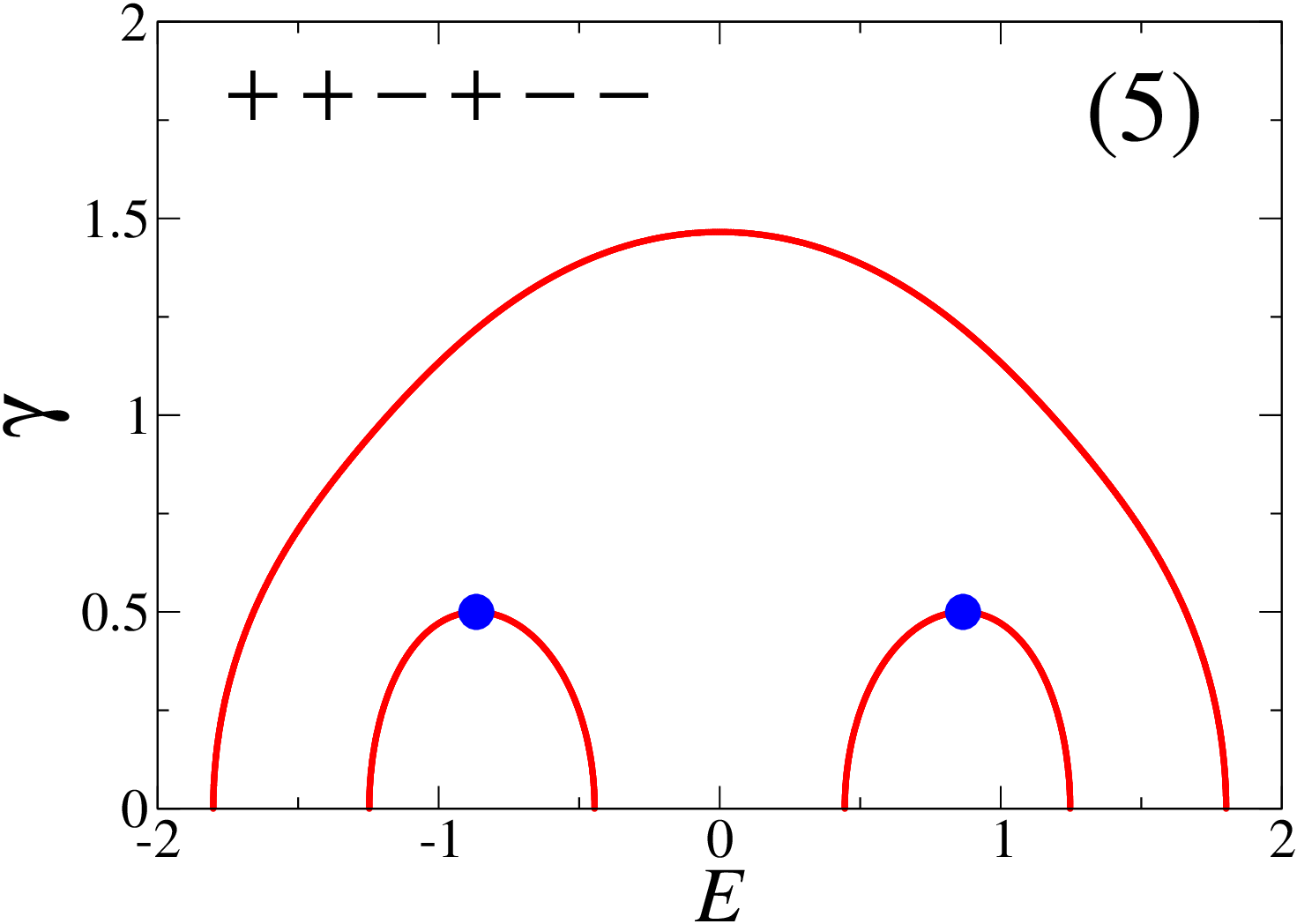}
\vskip 4pt
\includegraphics[angle=0,width=.4\linewidth,clip=true]{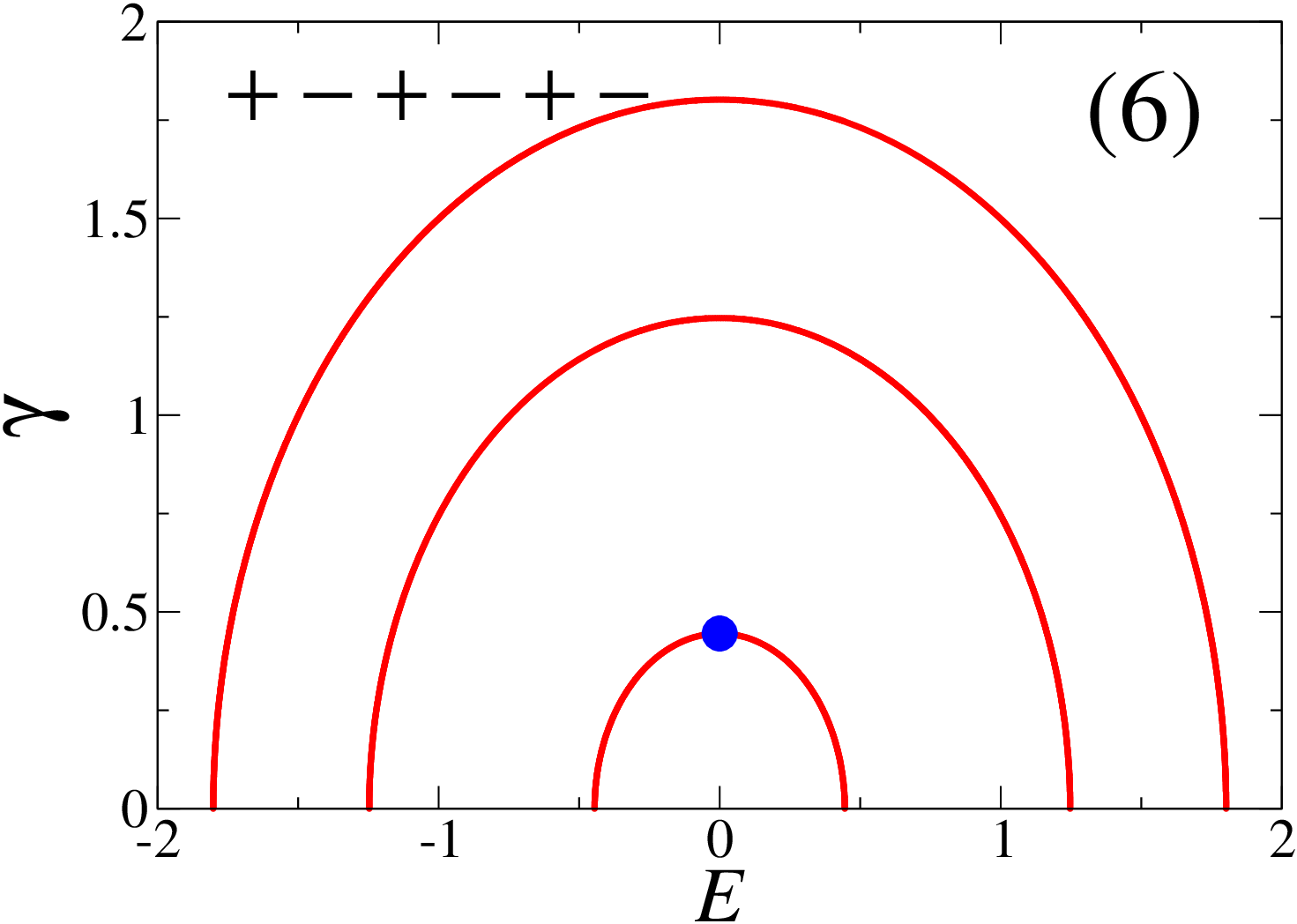}
\hskip 8pt
\includegraphics[angle=0,width=.4\linewidth,clip=true]{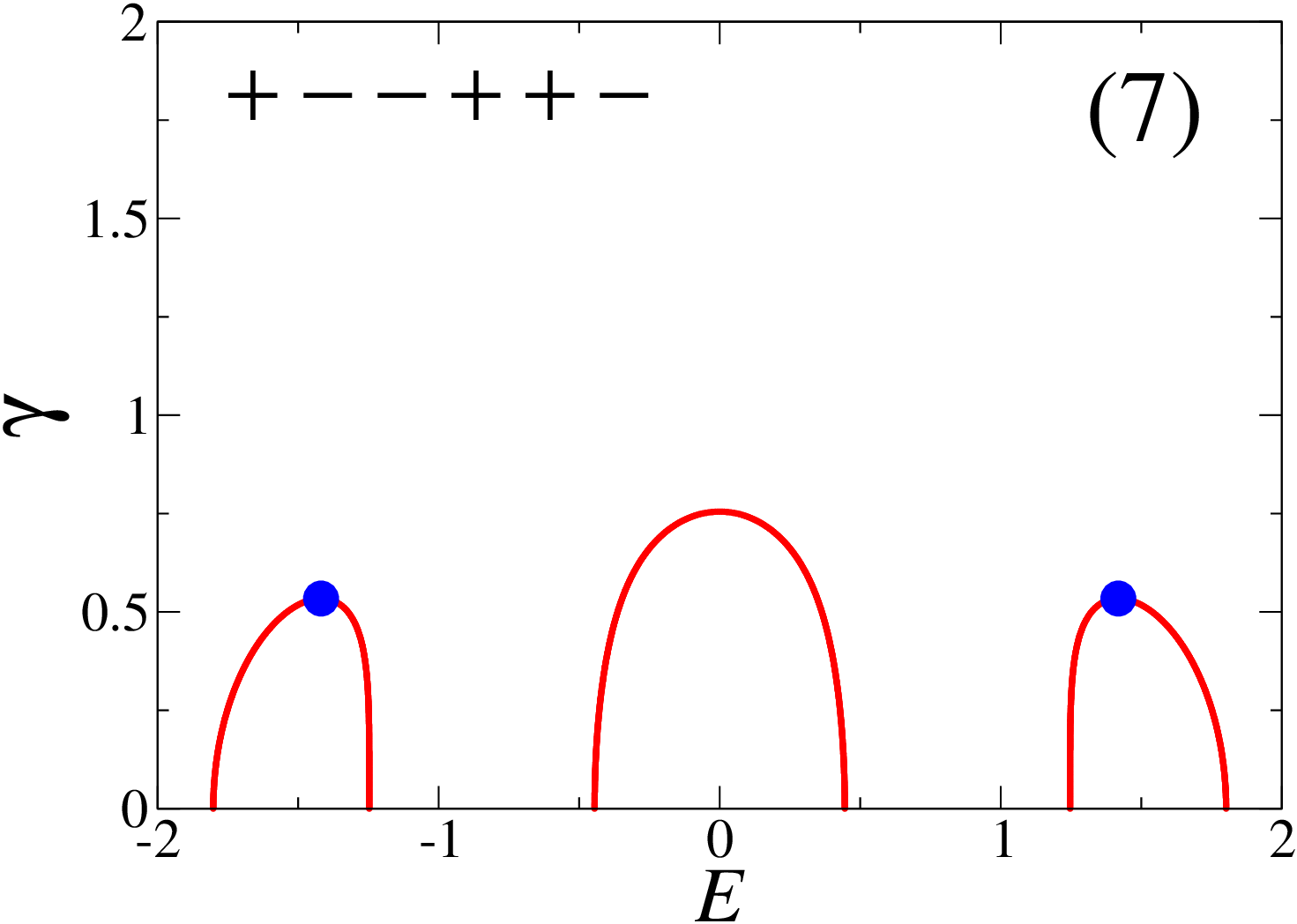}
\vskip 4pt
\caption{\small
Real eigenvalues of the seven shortest chains in the $E$--$\g$ plane (see text).
The full $\PT$-symmetric sequence is given in each case.
All plots are at the same scale in order to facilitate the comparison between them.
Blue symbols: thresholds $\g_c$ and corresponding degenerate eigenvalues.}
\label{eg}
\end{center}
\end{figure}

We begin our investigation with a detailed study of the shortest chains.
There are 1, 2 and 4 different half-sequences of lengths $M=1$, 2 and 3,
i.e., chain lengths $N=2M=2$, 4 and 6.
These sequences are successively investigated below.
Figure~\ref{eg} shows their real eigenvalues in the $E$--$\g$ plane for $\g\ge0$.
Blue symbols show the thresholds~$\g_c$ and the corresponding degenerate eigenvalues.
These plots exhibit a variety of topologically different patterns,
hinting at the complexity of the problem.

\smallskip
\noindent (1)
The half-sequence $(+)$ is the simplest of all.
It yields the full $\PT$-symmetric sequence $(+-)$.
Its characteristic polynomial is
\beq
P=E^2+\g^2-1.
\eeq
The energy eigenvalues $E=\pm\sqrt{1-\g^2}$ are real for $\g<1$ and imaginary for $\g>1$.
At the threshold
\beq
\g_c=1
\label{gshort1}
\eeq
there is a twofold degenerate eigenvalue at zero energy.

\noindent (2)
The half-sequence $(++)$ has characteristic polynomial
\beq
P=E^4+(2\g^2-3)E^2+\g^4+\g^2+1.
\eeq
At the threshold
\beq
\g_c=\frac{\sqrt{5}}{4}\approx0.559\,016
\label{gcst2}
\eeq
there are degenerate energy eigenvalues at $\pm E_c$, with
\beq
E_c=\frac{\sqrt{19}}{4}\approx1.089\,724.
\eeq

\noindent (3)
The half-sequence $(+-)$ has characteristic polynomial
\beq
P=E^4+(2\g^2-3)E^2+\g^4-3\g^2+1.
\eeq
At the threshold
\beq
\g_c=\frac{\sqrt{5}-1}{2}\approx0.618\,033
\label{gshort2}
\eeq
there is a twofold degenerate eigenvalue at zero energy.

\noindent (4)
The half-sequence $(+++)$ has characteristic polynomial
\beq
P=E^6+(3\g^2-5)E^4+(3\g^4-2\g^2+6)E^2+\g^6+3\g^4+2\g^2-1.
\eeq
At the threshold
\beq
\g_c\approx0.322\,142,
\eeq
obeying
\beq
2048\g_c^6+560\g_c^4+392\g_c^2-49=0,
\eeq
there are degenerate energy eigenvalues at $\pm E_c$, with
\beq
E_c\approx1.506\,895,
\eeq
obeying
\beq
2048E_c^6-7216E_c^4+6632E_c^2-1831=0.
\eeq

\noindent (5)
The half-sequence $(++-)$ has characteristic polynomial
\beq
P=E^6+(3\g^2-5)E^4+(3\g^4-6\g^2+6)E^2+\g^6-\g^4-2\g^2-1.
\eeq
At the threshold
\beq
\g_c=\frac{1}{2}
\eeq
there are degenerate energy eigenvalues at $\pm E_c$, with
\beq
E_c=\frac{\sqrt{3}}{2}\approx0.866\,025.
\eeq

\noindent (6)
The half-sequence $(+-+)$ has characteristic polynomial
\beq
P=E^6+(3\g^2-5)E^4+(3\g^4-10\g^2+6)E^2+\g^6-5\g^4+6\g^2-1.
\eeq
At the threshold
\beq
\g_c=2\sin\frac{\pi}{14}\approx0.445\,041
\label{gshort3}
\eeq
there is a twofold degenerate eigenvalue at zero energy.

\noindent (7)
The half-sequence $(+--)$ has characteristic polynomial
\beq
P=E^6+(3\g^2-5)E^4+(3\g^4-6\g^2+6)E^2+\g^6-\g^4+2\g^2-1.
\eeq
At the threshold
\beq
\g_c\approx0.534\,036,
\eeq
obeying
\beq
16\g_c^3+16\g_c^2-7=0,
\eeq
there are degenerate energy eigenvalues at $\pm E_c$, with
\beq
E_c\approx1.418\,436,
\eeq
obeying
\beq
256E_c^6-768E_c^4+576E_c^2-135=0.
\eeq

\subsection{Panoramic overview}
\label{pano}

\begin{figure}[!ht]
\begin{center}
\includegraphics[angle=0,width=.6\linewidth,clip=true]{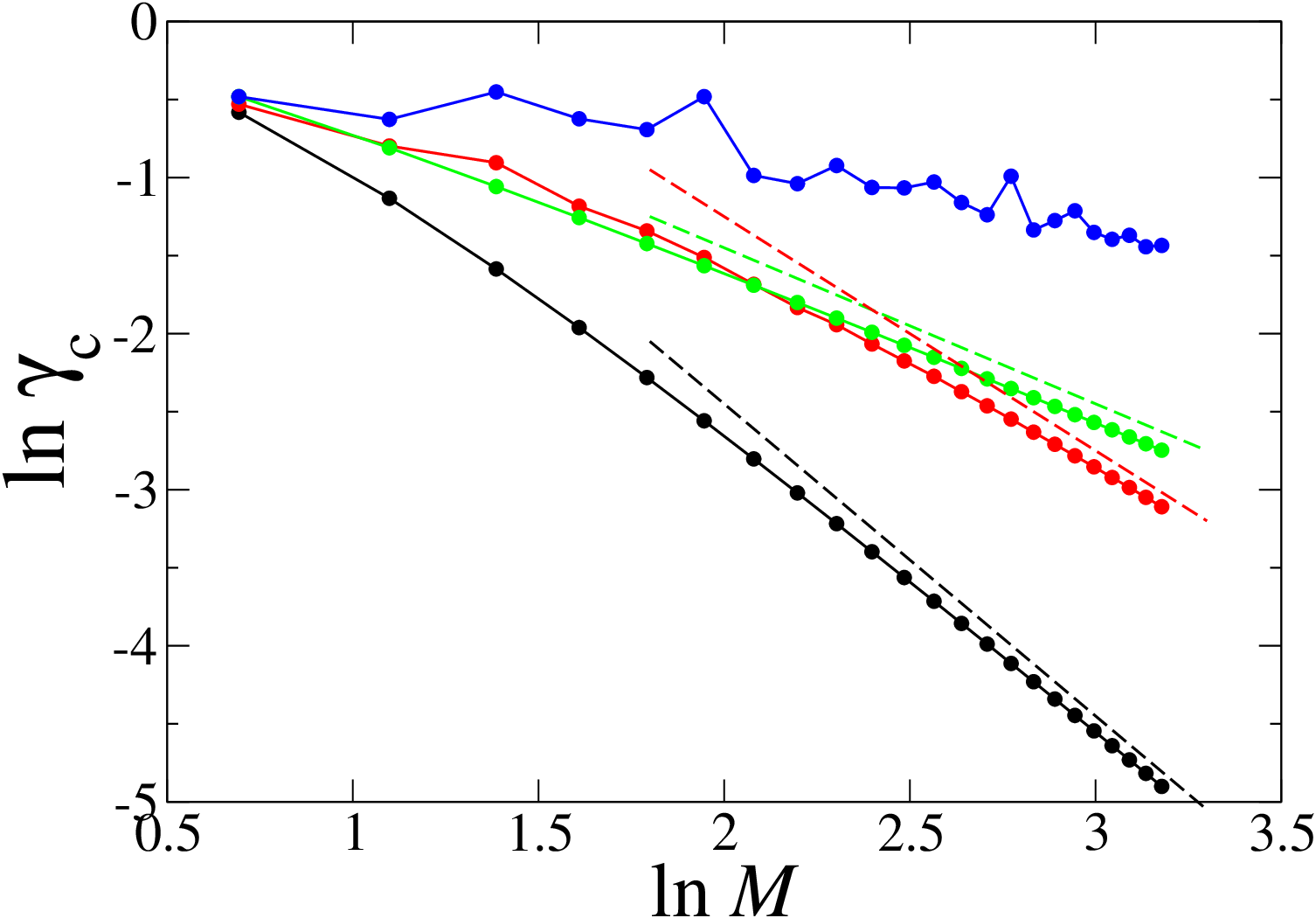}
\caption{\small
Log-log plot of various thresholds~$\g_c$
as a function of the length $M$ of the half-sequence, ranging from 2 to 24.
From bottom to top:
Black: smallest~$\g_c$ at fixed $M$,
reached for diblock chains.
Red: mean value of~$\g_c$ over all half-sequences of length $M$.
Green: $\g_c$ for alternating half-sequences.
Blue: largest~$\g_c$ at fixed $M$.
Dashed lines have the theoretical slopes $\beta=2$, $3/2$ and~1,
to be derived in the forthcoming sections.}
\label{panorama}
\end{center}
\end{figure}

We now turn to an overview of the behavior of the threshold $\g_c$ for $\PT$-symmetry breaking
for various kinds of configurations (periodic, random, etc.) of the optical potentials.
We have run a computer routine which enumerates
all the $2^{M-1}$ half-sequences of length $M$ starting with $\eps_1=+$,
up to a maximal length $M=24$, i.e., $N=48$,
and determines for each sequence accurate numerical values of the threshold
and of the corresponding degenerate energy eigenvalues.
The most notable emerging features are illustrated in Figure~\ref{panorama},
and will be studied in detail in the forthcoming sections.
In the regime of large chains,
thresholds fall off to zero, generically as a power law of the form
\beq
\g_c\sim\frac{1}{M^\beta}.
\label{betadef}
\eeq

Different classes of sequences yield different decay exponents $\beta$.
The smallest of all $\g_c$, decaying with exponent $\beta=2$,
is reached for the diblock chains having a constant half-sequence
(see black dataset and Section~\ref{diblock}).
The same scaling holds for all unbalanced periodic sequences
with a non-zero mean value (see~Section~\ref{uper}).
Alternating half-sequences,
as well as all balanced periodic ones with zero mean value,
have thresholds falling off with exponent $\beta=1$
(see green dataset and Section~\ref{alper}).
The mean value of~$\g_c$ over all half-sequences of given length
scales with exponent $\beta=3/2$.
The same scaling holds for typical random sequences,
i.e., for almost all half-sequences
(see red dataset and Section~\ref{ran}).
Finally, the largest threshold at fixed~$M$
has an irregular dependence on $M$ and presumably also scales asymptotically as $1/M$,
up to logarithmic corrections
(see blue dataset and Section~\ref{robust}).

\section{Diblock and other unbalanced periodic chains}
\label{diper}

\subsection{Diblock chains}
\label{diblock}

This section is devoted to diblock chains,
which are the most unbalanced of all $\PT$-symmetric chains.
On a diblock chain of of length $N=2M$,
the $M$ sites with gain are to the left,
and the $M$ sites with loss are to the right.
In other terms, the half-sequence is constant, i.e., $\eps_1=\dots=\eps_M=+$,
and so $\eps_{M+1}=\dots=\eps_{2M}=-$.
It is expected that the $\PT$-symmetric phase
is the most fragile on these diblock chains.
This is corroborated by the observation made in Section~\ref{pano}
that these chains yield the smallest threshold
at fixed chain length (see black dataset in Figure~\ref{panorama}).
The latter observation will now, in turn, be made quantitative by analytical means.

Instead of using a straightforward approach borrowed from elementary Quantum Mechanics,
i.e., looking for eigenfunctions of~(\ref{tbe}) as linear combinations of plane waves
with complex wavenumbers in each block,
we prefer to have recourse to the Riccati formalism introduced in Section~\ref{ric},
since the latter approach will prove more powerful in more complex situations.

The recursion~(\ref{recr}) obeyed by the Riccati variables does not depend explicitly
on $n$ and reads
\beq
R_n=2\cosh\mu-\frac{1}{R_{n-1}}\qquad(n=1,\dots,M),
\label{recd}
\eeq
where
\beq
\mu=\eta+\ii q
\eeq
is a complex variable such that
\beq
2\cosh\mu=E-\ii\g,
\eeq
i.e.,
\beq
E=2\cos q\;\cosh\eta,\qquad\g=-2\sin q\;\sinh\eta.
\eeq
The fixed points of~(\ref{recd}) are $R=\e^{\pm\mu}$.
Setting
\beq
Y_n=\frac{R_n-\e^{-\mu}}{R_n-\e^{\mu}},
\eeq
the recursion~(\ref{recd}) boils down to the multiplication by a constant, i.e.,
\beq
Y_n=\e^{2\mu}Y_{n-1}.
\eeq
The initial value $R_0=\infty$ yields $Y_0=1$, hence $Y_n=e^{2n\mu}$, and finally
\beq
R_n=\frac{\sinh(n+1)\mu}{\sinh n\mu}.
\label{rcst}
\eeq
The quantization condition~(\ref{qr}) therefore reads
\beq
\sinh(M+1)\mu\;\sinh(M+1)\mu^\star=\sinh M\mu\;\sinh M\mu^\star,
\eeq
i.e.,
\beq
\sin q\;\sin(2M+1)q+\sinh\eta\;\sinh(2M+1)\eta=0.
\label{qreal}
\eeq
The above equations are exact for any finite $M$.

We are mainly interested in the scaling behavior of $\g_c$ for long chains.
It turns out that the most unstable energy eigenvalues are close to the band edges
($E\to\pm2$).
Let us focus our attention onto the upper band edge ($E\to2$, i.e., $q\to0$)
and introduce the scaling variables
\beq
x=(2M+1)q,\qquad g=\frac{(2M+1)^2}{4}\,\g.
\eeq
We have then $(2M+1)\eta\approx(-2g/x)$,
and so the quantization condition~(\ref{qreal}) becomes
\beq
\sin x\approx-\frac{2g}{x^2}\sinh\frac{2g}{x},
\label{qsca}
\eeq
up to corrections of relative order $1/(2M+1)^2$.

\begin{figure}[!ht]
\begin{center}
\includegraphics[angle=0,width=.6\linewidth,clip=true]{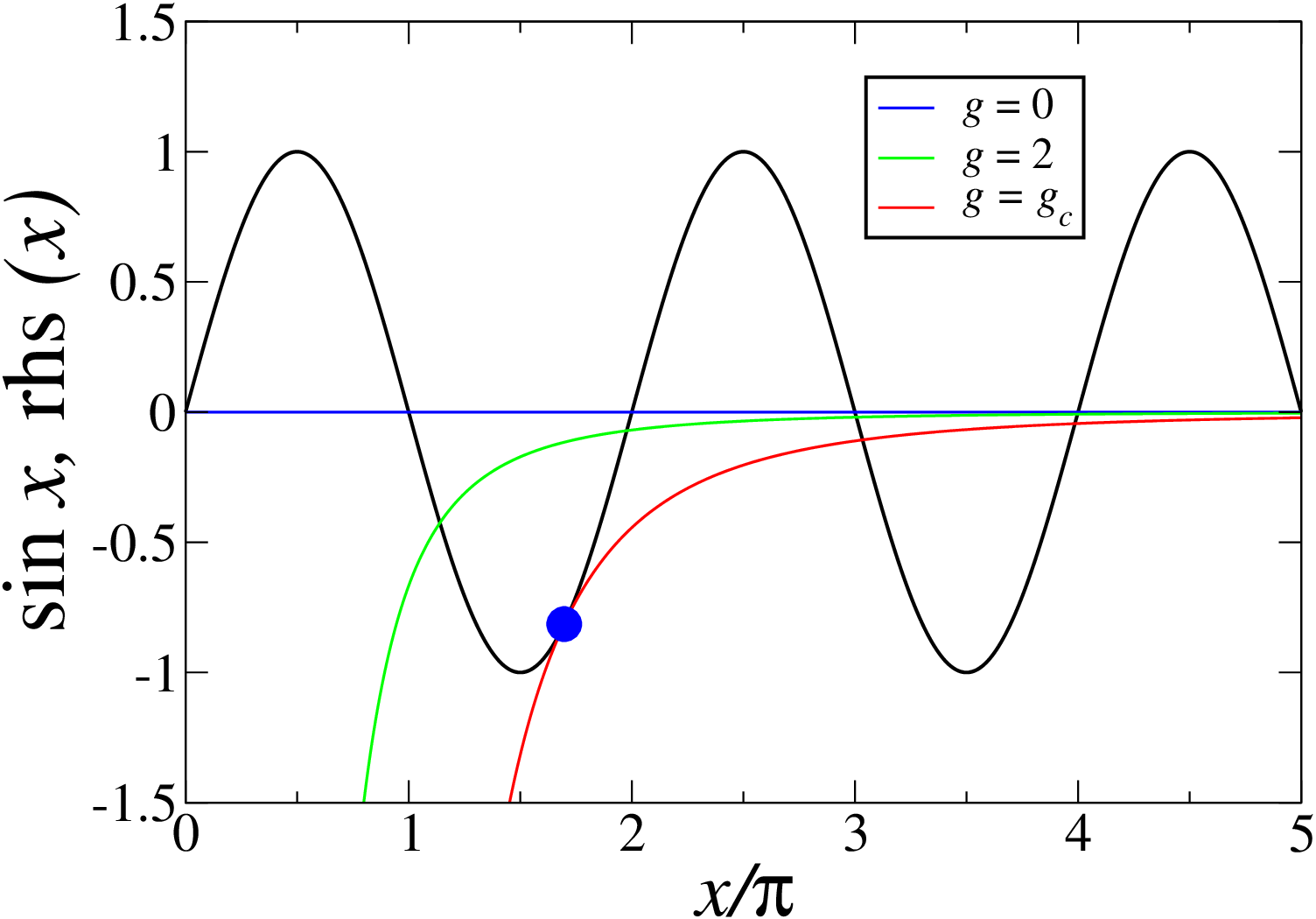}
\caption{\small
Graphical illustration of the threshold mechanism
for the rescaled quantization condition~(\ref{qsca})
near the upper edge of the spectrum of long diblock chains (see text).}
\label{merge}
\end{center}
\end{figure}

In the absence of optical potentials ($g=0$), the r.h.s.~of~(\ref{qsca}) vanishes.
We thus obtain
\beq
x\approx a\pi\qquad(a=1,2,\dots),
\label{api}
\eeq
in perfect agreement with~(\ref{ea}).
As the rescaled magnitude $g$ of gains and losses increases,
the rescaled spectrum is more and more distorted,
especially so for the very first values of $a$,
as shown in Figure~\ref{merge}.
The blue line $(g=0$) intersects the sine function at integer values of $x/\pi$
(see~(\ref{api})).
The green curve ($g=2$) yields a distorted albeit still entirely real spectrum.
The red one, corresponding~to
\beq
g_c\approx4.475\,308,
\label{gcnum}
\eeq
is tangent to the sine function at
\beq
x_c\approx5.331\,598
\eeq
(blue symbol),
where the lowest pair of eigenvalues ($a=1$ and 2) merges before becoming complex.
Other mergings between $a=3$ and 4, $a=5$ and 6, and so on,
take place at higher well-defined critical values of $g$.

In brief, for long diblock chains of length $N=2M$,
the threshold for $\PT$-symmetry breaking scales as
\beq
\g_c\approx\frac{4g_c}{(2M+1)^2}\sim\frac{g_c}{M^2}.
\label{gc}
\eeq
This provides a proof of the scaling law announced in Section~\ref{pano},
with exponent $\beta=2$,
together with a prediction for the amplitude $g_c$.
The first two pairs of degenerate eigenvalues sit
very near the band edges, at $\pm E_c$, with
\beq
2-E_c\approx q^2-\eta^2\approx\frac{4\sigma_c}{(2M+1)^2}\sim\frac{\sigma_c}{M^2},
\label{edge}
\eeq
with
\beq
\sigma_c=\frac{x_c^2}{4}-\frac{g_c^2}{x_c^2}\approx6.401\,903.
\eeq
Both estimates~(\ref{gc}) and~(\ref{edge}) hold up to corrections of relative order $1/(2M+1)^2$.

\begin{figure}[!ht]
\begin{center}
\includegraphics[angle=0,width=.6\linewidth,clip=true]{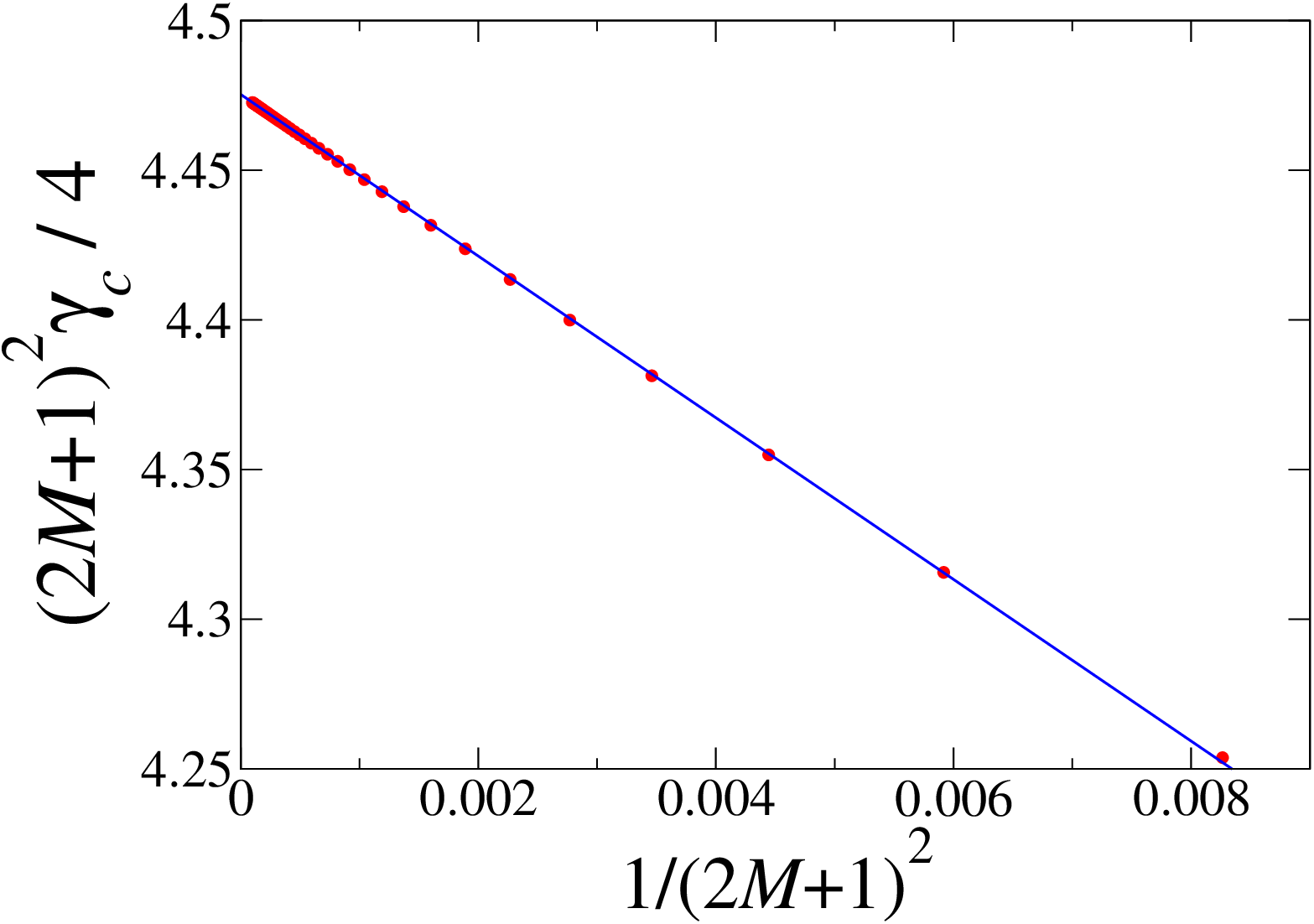}
\caption{\small
Red symbols:
combination $(2M+1)^2\g_c/4$ plotted against $1/(2M+1)^2$,
where $\g_c$ is the threshold of diblock chains.
The blue line has the theoretical intercept $g_c$ given in~(\ref{gcnum}).}
\label{gc1}
\end{center}
\end{figure}

Figure~\ref{gc1} provides a quantitative check of the prediction~(\ref{gc}),
including the order of magnitude of the corrections.
The combination $(2M+1)^2\g_c/4$ is plotted against $1/(2M+1)^2$, up to $M=50$.
The blue line has the theoretical intercept $g_c$ given in~(\ref{gcnum}).

\subsection{Unbalanced periodic chains}
\label{uper}

We now turn to unbalanced periodic chains.
A $\PT$-symmetric chain is said to be periodic
if the half-sequence $\eps_1,\dots,\eps_M$ is periodic,
i.e., $\eps_n=\eps_{n+L}$ for some period $L\ge1$ and all $n=1,\dots,M-L$.
The motif $(\eps_1,\dots\eps_L)$ is called the unit cell of the periodic sequence.
In this section we make the assumption that the periodic sequence
is unbalanced, i.e., its mean value $\bar\eps$
is non-zero.\footnote{Throughout this paper a bar denotes a spatial average.
If the sequence is disordered, e.g.~with independent entries,
this amounts to averaging over disorder.}
In the present case, we have
\beq
\bar\eps=\frac{1}{L}\sum_{n=1}^L\eps_n=\frac{L_+-L_-}{L},
\label{barepsdef}
\eeq
where $L_+$ and $L_-$ are the numbers of plus and minus signs
(i.e., of sites with gain and loss) in the unit cell, with $L_++L_-=L$.

For the time being, let us follow a heuristic line of thought.
If $\bar\eps$ is positive,
on a spatial scale much larger than the period $L$,
the left half of the chain has a roughly homogeneous gain,
whereas the right half has a roughly homogeneous loss, and vice versa if $\bar\eps$ is negative.
Any unbalanced periodic sequence therefore amounts to an effective diblock sequence,
with $\g$ being replaced by the effective value $\g_{\rm eff}=\g\bar\eps$.
This approach predicts that the threshold
of a long unbalanced periodic chain
is asymptotically given by the product of the expression~(\ref{gc})
for diblock chains and of an amplification factor
\beq
f=\frac{1}{\abs{\bar\eps}}=\frac{L}{\abs{L_+-L_-}},
\eeq
depending only on the contents of the unit cell of the sequence.
This reads
\beq
\g_c\approx\frac{4fg_c}{(2M+1)^2}\sim\frac{fg_c}{M^2}.
\label{gcup}
\eeq
Consistently,
diblock chains themselves can be viewed as unbalanced periodic ones with $L=L_+=1$,
and so $f=1$.
Allowing larger and larger periods,
the amplification factor $f$ may become
either arbitrarily close to unity or arbitrarily large:
if $L\ge3$ and $L_\pm=1$, we have $f=L/(L-2)$;
if $L$ is odd and $L_\pm=(L-1)/2$, we have $f=L$.

The result~(\ref{gcup}) turns out to be asymptotically correct,
in spite of its heuristic derivation.
This will be shown in a broader setting in Section~\ref{cric},
by means of a continuum Riccati approach.

\begin{figure}[!ht]
\begin{center}
\includegraphics[angle=0,width=.6\linewidth,clip=true]{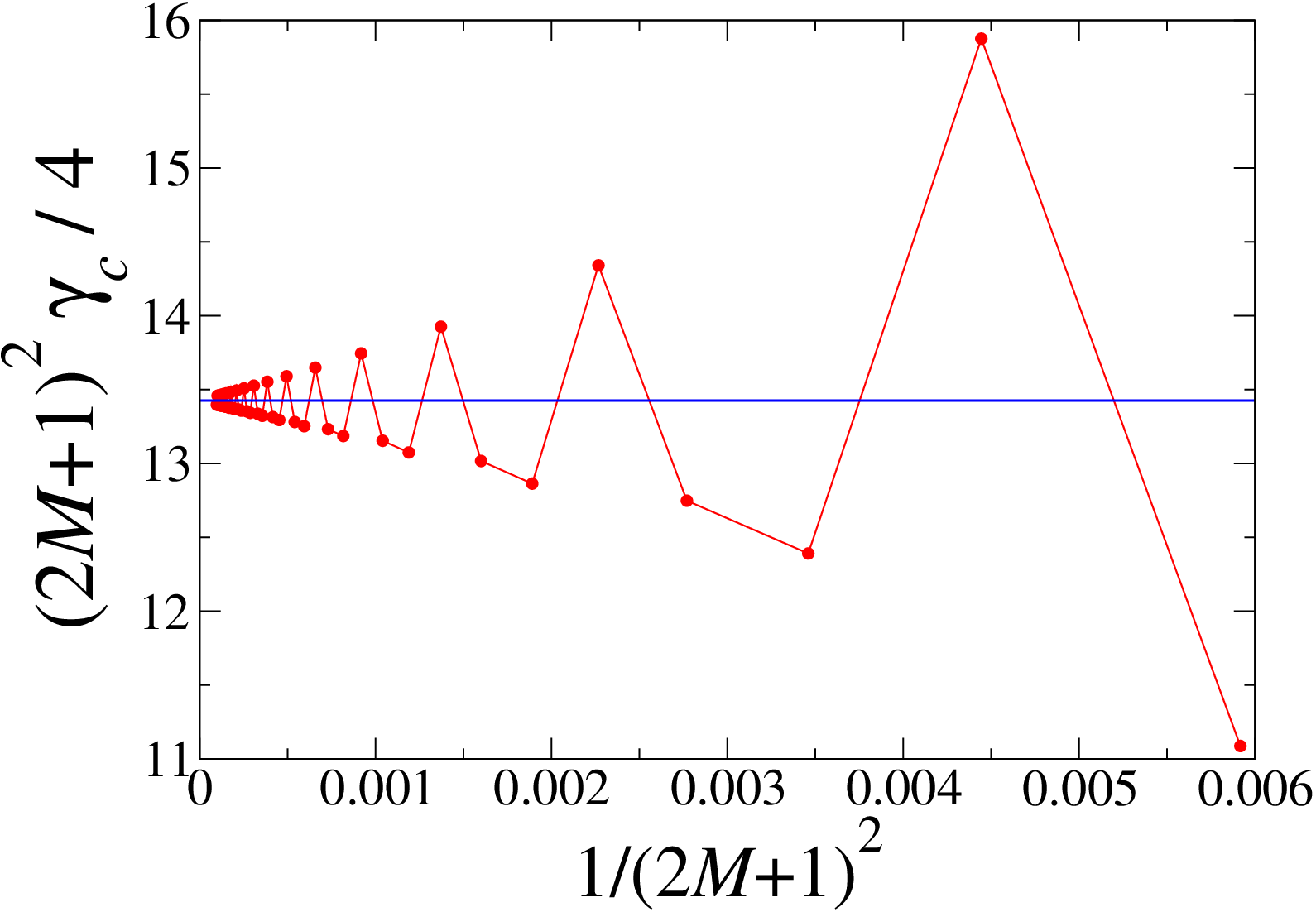}
\caption{\small
Red symbols:
combination $(2M+1)^2\g_c/4$ plotted against $1/(2M+1)^2$,
where $\g_c$ is the threshold of unbalanced chains with unit cell $(++-)$.
Horizontal blue line: predicted limit value $3g_c\approx13.425$.}
\label{gc3}
\end{center}
\end{figure}

The non-trivial unbalanced chain with shortest period
has $L=3$ and unit cell $(++-)$, hence $f=3$.
Figure~\ref{gc3} provides a quantitative check of the prediction~(\ref{gcup})
on this example.
Here again, the combination $(2M+1)^2\g_c/4$ is plotted against $1/(2M+1)^2$, up to $M=50$.
Data for finite chains converge to the predicted limit value $3g_c\approx13.425$
(horizontal blue line), oscillating with the period $L=3$
of the underlying sequence of gains and losses.
Oscillations are damped linearly on the chosen scale.

\section{Alternating and other balanced periodic chains}
\label{alper}

\subsection{Alternating chains}
\label{alt}

In this section we consider alternating chains,
where sites with gains and losses alternate along the chain,
so that $\eps_n=(-1)^n$.
If the chain length $N=2M$ is even,
this configuration of optical potentials obeys $\PT$-symmetry.
Alternating chains are periodic with period $L=2$ and balanced,
in the sense that $\bar\eps=0$,
so that the analysis of Section~\ref{uper} does not apply.

Our goal is to derive the exact threshold $\g_c$ for any finite chain length $N=2M$
(see~(\ref{gcexact2})).
The analysis again relies on the Riccati formalism.
The recursion~(\ref{recr}) takes two different forms according to the parity of $n$:
\beq
R_{2p+1}=E+\ii\g-\frac{1}{R_{2p}},\qquad
R_{2p+2}=E-\ii\g-\frac{1}{R_{2p+1}}.
\label{rec1}
\eeq
These two equations can be combined, yielding
\beq
R_{2p+2}=\frac{(E^2+\g^2-1)R_{2p}-E+\ii\g}{(E+\ii\g)R_{2p}-1}.
\label{rec2}
\eeq
Introducing the parametrization
\beq
E+\ii\g=2\cos q\,\e^{\ii\theta},
\eeq
i.e.,
\beq
E=2\cos q\;\cos\theta,\qquad\g=2\cos q\;\sin\theta,
\eeq
and so
\beq
E^2+\g^2=4\cos^2q,
\label{dis2}
\eeq
the fixed points of~(\ref{rec2}) are $R=\e^{-\ii(\theta\pm q)}$.
Setting
\beq
Y_n=\frac{R_n-\e^{-\ii(\theta+q)}}{R_n-\e^{-\ii(\theta-q)}},
\eeq
the recursion~(\ref{rec2}) boils down to
\beq
Y_{2p+2}=\e^{4\ii q}Y_{2p}.
\label{recy}
\eeq
The initial value $R_0=\infty$ yields $Y_0=1$, hence $Y_{2p}=e^{4\ii pq}$, and finally
\beq
R_{2p}=\e^{-\ii\theta}\;\frac{\sin(2p+1)q}{\sin 2pq},\qquad
R_{2p+1}=\e^{\ii\theta}\;\frac{\sin(2p+2)q}{\sin(2p+1)q}.
\label{ralt}
\eeq

The quantization condition~(\ref{qr}) reads
\beq
\sin(2M+1)q=0,
\eeq
irrespective of $\theta$ and of the parity of $M$.
This equation is considerably simpler than its counterpart~(\ref{qreal}) for diblock chains.
It yields the explicit result
\beq
q=\frac{k\pi}{2M+1}\qquad(k=1,\dots,M).
\label{qq}
\eeq

For each quantized value of $q$, labelled by the integer~$k$,
a pair of eigenvalues merges at zero energy as $\theta\to\pi/2$,
whereas $\g$ reaches the critical value $\g_k=2\cos q_k$.
The pattern of real eigenvalues in the $E$--$\g$ plane
therefore has the form of a rainbow of $M$ nested arches
(see panels (1), (3) and (6) of Figure~\ref{eg}).
The first merging event corresponds to the smallest critical value, i.e., $k=M$.
We thus obtain the exact expression
\beq
\g_c=2\sin\frac{\pi}{2(2M+1)}
\label{gcexact2}
\eeq
for the threshold of alternating $\PT$-symmetric chains of any finite length.
We recover known values of $\g_c$ for the first values of $M$, namely
$\g_c=2\sin\pi/6=1$ for $M=1$ (see~(\ref{gshort1})),
$\g_c=2\sin\pi/10=(\sqrt5-1)/2$ for $M=2$ (see~(\ref{gshort2})),
and $\g_c=2\sin\pi/14$ for $M=3$ (see~(\ref{gshort3})).
For long alternating chains,
the threshold falls off as
\beq
\g_c\approx\frac{\pi}{2M}.
\label{gcalt}
\eeq
This law of decay with exponent $\beta=1$ was announced in Section~\ref{pano}.

\subsection{One example with period four}
\label{perfour}

We now turn to an example of balanced periodic chains
where pairs of sites with gains and losses form an alternating pattern.
These chains have period $L=4$ and unit cell $(++--)$.
The analysis again relies on the Riccati formalism.
The recursion~(\ref{recr}) reads
\beqa
R_{4p+1}=E-\ii\g-\frac{1}{R_{4p}},\qquad
R_{4p+2}=E-\ii\g-\frac{1}{R_{4p+1}},
\nonumber\\
R_{4p+3}=E+\ii\g-\frac{1}{R_{4p+2}},\qquad
R_{4p+4}=E+\ii\g-\frac{1}{R_{4p+3}}.
\eeqa
These four equations can be combined into a recursion
expressing $R_{4p+4}$ as a function of $R_{4p}$,
whose expression is more intricate than~(\ref{rec2}).
Anticipating that the threshold
corresponds to a merging of eigenvalues around $E_c=\sqrt2$
(see the discussion below~(\ref{eres})),
we set $E=\sqrt2+a$, and simplify the recursion
in the scaling regime where both $a$ and $\g$ are small.
We thus obtain
\beq
R_{4p+4}=\frac{(1-2\sqrt2\,a)R_{4p}+2\ii\g+4a}{1+2\sqrt2\,a+(2\ii\g-4a)R_{4p}}.
\label{rec4}
\eeq
Introducing parameters $w$, $\phi$ and $\theta$ such that
\beq
a=\half w\cos\phi,\qquad\g=w\sin\phi,\qquad\cos\theta=\frac{\cos\phi}{\sqrt2},
\eeq
with $0<\phi<\pi$ and $\pi/4<\theta<3\pi/4$,
the fixed points of~(\ref{rec4}) are $R=\e^{\ii(\phi\pm\theta)}$.
Setting
\beq
Y_{4p}=\frac{R_{4p}-\e^{\ii(\phi-\theta)}}{R_{4p}-\e^{\ii(\phi+\theta)}},
\eeq
the recursion~(\ref{rec4}) boils down to
\beq
Y_{4p+4}=\e^{-2\ii\lambda}Y_{4p},
\eeq
with
\beq
\tan\lambda=2w\sin\theta.
\eeq
The initial value $R_0=\infty$ yields $Y_0=1$, hence $Y_{4p}=e^{-2\ii p\lambda}$, and finally
\beq
R_{4p}=\e^{\ii\phi}\;\frac{\sin(p\lambda-\theta)}{\sin p\lambda},\qquad
R_{4p+1}=\sqrt2-\e^{-\ii\phi}\;\frac{\sin p\lambda}{\sin(p\lambda-\theta)}.
\eeq

For $M=4p$ (and also for $M=4p+2$),
the quantization condition~(\ref{qr}) reads
\beq
\sin(p\lambda-\theta)=\pm\sin p\lambda,
\eeq
yielding
\beq
\lambda=\frac{\theta+k\pi}{2p},
\eeq
with $k$ integer.
The threshold is expected to correspond to the smallest value of $\lambda$, i.e., $k=0$.
It can be shown that the rescaled combinations
\beq
G=M\g,\qquad x=Ma=M(E-\sqrt2),
\label{rescacom}
\eeq
asymptotically only depend on $\theta$, according to
\beq
G=\frac{\theta\sqrt{-\cos 2\theta}}{\sin\theta},\qquad
x=\frac{\theta\cot\theta}{\sqrt2}.
\label{gxcurve}
\eeq
The threshold $\g_c$ corresponds to the point in the $x$--$G$ plane
where the curve given parametrically by~(\ref{gxcurve})
has a horizontal tangent, as in~Figure~\ref{eg},
i.e., to the maximum of the function $G(\theta)$.
As $\theta$ is varied between $\pi/4$ and $3\pi/4$,
$G$ reaches its maximum~$G_1$ for $\theta_1$ such that $\theta_1=\cos2\theta_1\tan\theta_1$,
i.e., $\theta_1\approx1.937\,183$, where we have
\beq
G_1\approx1.788\,897,\qquad x_1\approx-0.525\,607.
\label{evenc}
\eeq

For $M=4p+1$ (and also for $M=4p+3$),
the quantization condition~(\ref{qr})~reads
\beq
\sin^2(p\lambda-\theta)+\sin^2p\lambda
-4\cos\theta\;\sin p\lambda\;\sin(p\lambda-\theta)=0,
\eeq
whose smallest solution is
\beq
\lambda=\frac{\theta+\alpha}{2p},
\eeq
with
\beq
\cos\alpha=\frac{\cos2\theta}{\cos\theta}
\eeq
and $0<\alpha<\pi$.
The quantities introduced in~(\ref{rescacom}) now read
\beq
G=\frac{(\theta+\alpha)\sqrt{-\cos 2\theta}}{\sin\theta},\qquad
x=\frac{(\theta+\alpha)\cot\theta}{\sqrt2}.
\eeq
The threshold is obtained by the maximization procedure applied to~(\ref{gxcurve}).
As $\theta$ is varied between $\pi/4$ and $3\pi/4$,
$G$ reaches its maximum $G_2$
for $\theta_2\approx2.179\,171$, where we have
\beq
G_2\approx2.223\,114,\qquad x_2\approx-1.525\,862.
\label{oddc}
\eeq

To sum up,
for large chains with even $M$,
the threshold scales as
\beq
\g_c\approx\frac{G_1}{M},
\eeq
and corresponds to two pairs of degenerate eigenvalues at $\pm E_c$, with
\beq
E_c\approx\sqrt2+\frac{x_1}{M},
\eeq
where $G_1$ and $x_1$ are given in~(\ref{evenc}).
For large chains with odd $M$,
the threshold scales~as
\beq
\g_c\approx\frac{G_2}{M},
\eeq
and corresponds to two pairs of degenerate eigenvalues at $\pm E_c$, with
\beq
E_c\approx\sqrt2+\frac{x_2}{M},
\eeq
where $G_2$ and $x_2$ are given in~(\ref{oddc}).

\begin{figure}[!ht]
\begin{center}
\includegraphics[angle=0,width=.6\linewidth,clip=true]{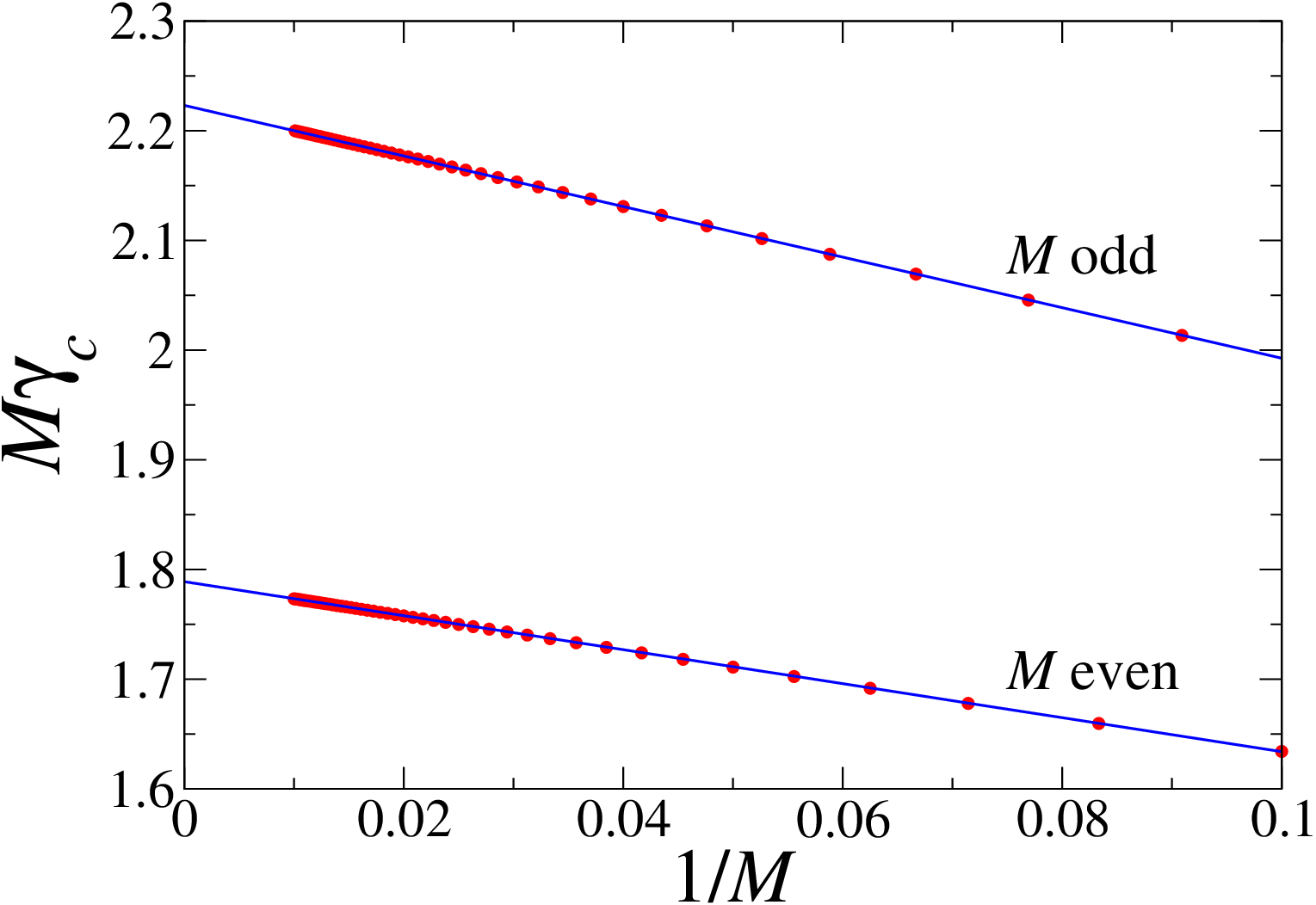}
\caption{\small
Red symbols:
product $M\g_c$ for chains with unit cell $(++--)$,
plotted against $1/M$ up to $M=100$.
The upper (resp.~lower) dataset corresponds to odd (resp.~even) values of $M$.
The blue lines have the theoretical intercepts $G_1$ and $G_2$ (see~(\ref{evenc}),~(\ref{oddc})).}
\label{gc4}
\end{center}
\end{figure}

Figure~\ref{gc4} provides a quantitative check of the above predictions.
The product $M\g_c$ is plotted against $1/M$, up to $M=100$.
The upper (resp.~lower) dataset corresponds to odd (resp.~even) values of $M$.
The blue lines have the theoretical intercepts $G_1$ and~$G_2$ (see~(\ref{evenc}),~(\ref{oddc})).

\subsection{Balanced periodic chains}
\label{bper}

Let us now turn to balanced periodic chains
with even period $L$ and $\bar\eps=0$ (see~(\ref{barepsdef})).
Only a few general features can be stated in the present situation.
This is in strong contrast with unbalanced chains,
investigated in Section~\ref{uper},
where the quantitative prediction~(\ref{gcup}) has been derived in full generality.

The Fourier transform of a periodic chain with period $L$
is supported by all integer multiples of the reciprocal period $2\pi/L$.
This singles out $L-1$ resonant energies
inside the band of the free tight-binding chain, i.e.,
\beq
E_a=2\cos\frac{a\pi}{L}\qquad(a=1,\dots,L-1).
\label{eres}
\eeq
For long balanced periodic chains $(M\gg1)$,
it is expected that the energy $E_c$ at which the first merging of eigenvalues
takes place approaches one of those resonant energies.
The convention $E_c\ge0$ selects $a=1,\dots,L/2$, the last case yielding $E_c=0$.
For the alternating chain (see Section~\ref{alt}),
$L=2$ and $a=1$ correctly predict $E_c=0$.
For the chain studied in Section~\ref{perfour},
$L=4$ and $a=1$ correctly predict $E_c=\sqrt2$,
since $a=2$ is ruled out by symmetry.
For chains with larger periods $L$,
there is no systematic way of predicting the resonant energy $E_c$.
There are actually examples where $E_c$ keeps varying periodically with $M$ mod $L$.

The threshold for $\PT$-symmetry breaking is expected to decay as
\beq
\g_c\approx\frac{G_c}{M}
\label{algal}
\eeq
for all balanced periodic chains.
For the alternating chain (see Section~\ref{alt}),
we have $G_c=\pi/2$ (see~(\ref{gcalt})).
For the chain with period four studied in Section~\ref{perfour},~$G_c$ alternates
between the two numbers $G_1$ and $G_2$,
according to the parity of $M$ (see~(\ref{evenc}),~(\ref{oddc})).
For chains with larger periods,
the amplitude $G_c$ is generically expected to vary periodically with $M$ mod $L$.
Many questions remain open,
including whether the value $\pi/2$ obtained for alternating chains is a lower bound for $G_c$
and whether $G_c$ may become arbitrarily large.

\section{Random chains}
\label{ran}

This section is devoted to random $\PT$-symmetric chains,
where each symbol of the half-sequence $(\eps_1,\dots,\eps_M)$
is taken to be $\eps_n=\pm$ with equal probabilities.
Equivalently, the $2^M$ different half-sequences are considered as equally probable.
The threshold $\g_c$ for $\PT$-symmetry breaking is now a random variable,
as it depends on the chain under scrutiny.
This situation has been studied by numerical means in~\cite{ref5}.

\subsection{Continuum Riccati formalism}
\label{cric}

The forthcoming analysis again relies on the Riccati formalism introduced in Section~\ref{ric}.
We are mainly interested in the behavior of the threshold $\g_c$ for long random chains.
In this regime, $\g_c$ is typically small,
whereas the first mergings concern pairs
of eigenvalues close to the band edges ($E\to\pm2$).

Let us focus for definiteness our attention onto the upper band edge ($E\to2$).
In the vicinity of the latter, setting $E=2-q^2$,
in the presence of small optical potentials $\ii\g_n=\ii\g\eps_n$,
it is advantageous to consider the small variable
\beq
Z_n=R_n-1,
\eeq
in terms of which the recursion~(\ref{recr}) reads
\beq
Z_n=-q^2-\ii\g_n+\frac{Z_{n-1}}{1+Z_{n-1}}
=-q^2-\ii\g_n+Z_{n-1}-Z_{n-1}^2+\cdots
\eeq
This expansion suggests to use a continuum formalism,
considering $Z(n)$ as a continuous function of the real variable $n$,
obeying the Riccati differential equation
\beq
\frac{\dd Z}{\dd n}=-q^2-\ii\g(n)-Z^2,
\label{riceq}
\eeq
with initial condition
\beq
Z(0)=\infty.
\label{zinit}
\eeq
In spite of this, the function $Z(n)$ is small all over the relevant range.
In particular, if the optical potential is constant,
setting $\mu^2=-q^2-\ii\g$,
the solution to~(\ref{riceq}) reads
\beq
Z=\mu\cotanh n\mu.
\eeq
To leading order as $\mu\to0$,
for fixed values of the product $n\mu$,
this is in agreement with~(\ref{rcst}),
which can be recast as
\beq
R_n=\cosh\mu+\sinh\mu\cotanh n\mu=1+\mu\cotanh n\mu+\half\mu^2+\cdots
\eeq

Before we pursue with random chains,
let us come back for a while to unbalanced periodic chains,
investigated in Section~\ref{uper}.
If the sequence $\eps_n$ is periodic with period~$L$ and has a non-zero mean value $\bar\eps$,
the continuous function $\g(n)$ has the same period and mean value $\bar\g=\g\bar\eps$.
In the situation of interest, namely sequences of length $M\gg L$,
the function $Z(n)$ varies typically on a scale $M$.
It is therefore slowly varying at the scale of one period,
so that it is legitimate to replace $\g(n)$ by its mean value $\bar\g$.
This puts the heuristic line of thought of Section~\ref{uper} on a firmer basis.

\subsection{Predictions for random chains}
\label{pran}

For long random $\PT$-symmetric chains,
the function $\g(n)$ becomes in the continuum limit
a Gaussian white noise with covariance
\beq
\bar{\g(m)\g(n)}=\g^2\delta(m-n).
\label{gcovar}
\eeq
The continuum Riccati equation~(\ref{riceq}) for the function $Z(n)$
has to be integrated from $n=0$, with initial condition~(\ref{zinit}),
to $n=M$, the length of the half-sequence,
where the quantization condition~(\ref{qr}) translates to
\beq
\Re Z(M)=0.
\label{zfinal}
\eeq

The Riccati differential equation driven by a real Gaussian white noise, i.e.,
\beq
\frac{\dd Z}{\dd x}=-q^2-V(x)-Z^2,
\label{ricr}
\eeq
with
\beq
\bar{V(x)V(y)}=w^2\delta(x-y),
\label{vcovar}
\eeq
has a long history, dating back to pioneering studies in the 1960s
of Anderson localization in a one-dimensional white-noise potential~\cite{FL,Hal}.
The Lyapunov exponent
\beq
\Gamma=\bar{Z(\infty)},
\label{gammadef}
\eeq
representing the inverse localization length of the model on an infinite line,
obeys a scaling law of the form
\beq
\Gamma\approx\left(\half w^2\right)^{1/3}\,F(x),\qquad
x=\left(\half w^2\right)^{-2/3}\,q^2,
\label{lya}
\eeq
in the low-energy scaling regime where $q$ and $w$ are simultaneously small.
The scaling function $F(x)$ is known in terms of Airy functions:
\beq
F(x)=\frac{\Ai(x)\Ai'(x)+\Bi(x)\Bi'(x)}{\Ai(x)^2+\Bi(x)^2}.
\label{airy}
\eeq
The result~(\ref{lya}) is commonly referred to as anomalous band-edge scaling.
It has been rederived several times,
both in continuum and in lattice models~\cite{DG,IRT,CLTT}.
Several works have been devoted to the fluctuations of the process $Z(x)$
and of closely related ones around their stationary mean value~\cite{ST,RT,FDRT}.
The mathematics involved there is quite intricate.

Fortunately enough, predictions of interest concerning the statistics of the threshold $\g_c$
of long random chains can be derived from the sole consideration of scale invariance.
The continuous Riccati equations~(\ref{riceq}),~(\ref{ricr})
driven by noises obeying~(\ref{gcovar}),~(\ref{vcovar})
are invariant under the scaling transformation
\beq
(n,x)\to\frac{1}{\lambda}\,(n,x),\quad
(Z,q)\to\lambda(Z,q),\quad
(\g,w)\to\lambda^{3/2}(\g,w).
\label{rg}
\eeq

For the problem of Anderson localization,
the scale invariance of the Riccati equation~(\ref{ricr})
dictates the scaling form~(\ref{lya}) of the Lyapunov exponent.
Of course, the determination of the exact form~(\ref{airy})
of the scaling function $F(x)$ requires a more advanced analysis.

In the present context,
the full scale invariance of the Riccati equation~(\ref{riceq}),
including its
boundary conditions~(\ref{zinit}) at $n=0$ and~(\ref{zfinal}) at $n=M$,
implies that the mean value of the threshold of long random chains
scales as
\beq
\bar{\g_c}\approx\frac{A}{M^{3/2}},
\label{gmoy}
\eeq
with a decay exponent $\beta=3/2$,
as announced in Section~\ref{pano},
and that the reduced random variable
\beq
\xi=\frac{\g_c}{\,\bar{\g_c}\,}
\label{xidef}
\eeq
has a well-defined limiting distribution with density $f(\xi)$, such that $\bar\xi=1$.
It should be clear from the above reasoning that the latter distribution is universal,
in the sense that it would hold for any symmetric, i.e., even distribution
of gain/loss optical potentials~$\g_n$ with finite variance.
The determination of the amplitude $A$ and of the distribution~$f(\xi)$
by analytical means remains a difficult open problem.
We therefore have to rely on numerical analysis.

\begin{figure}[!ht]
\begin{center}
\includegraphics[angle=0,width=.6\linewidth,clip=true]{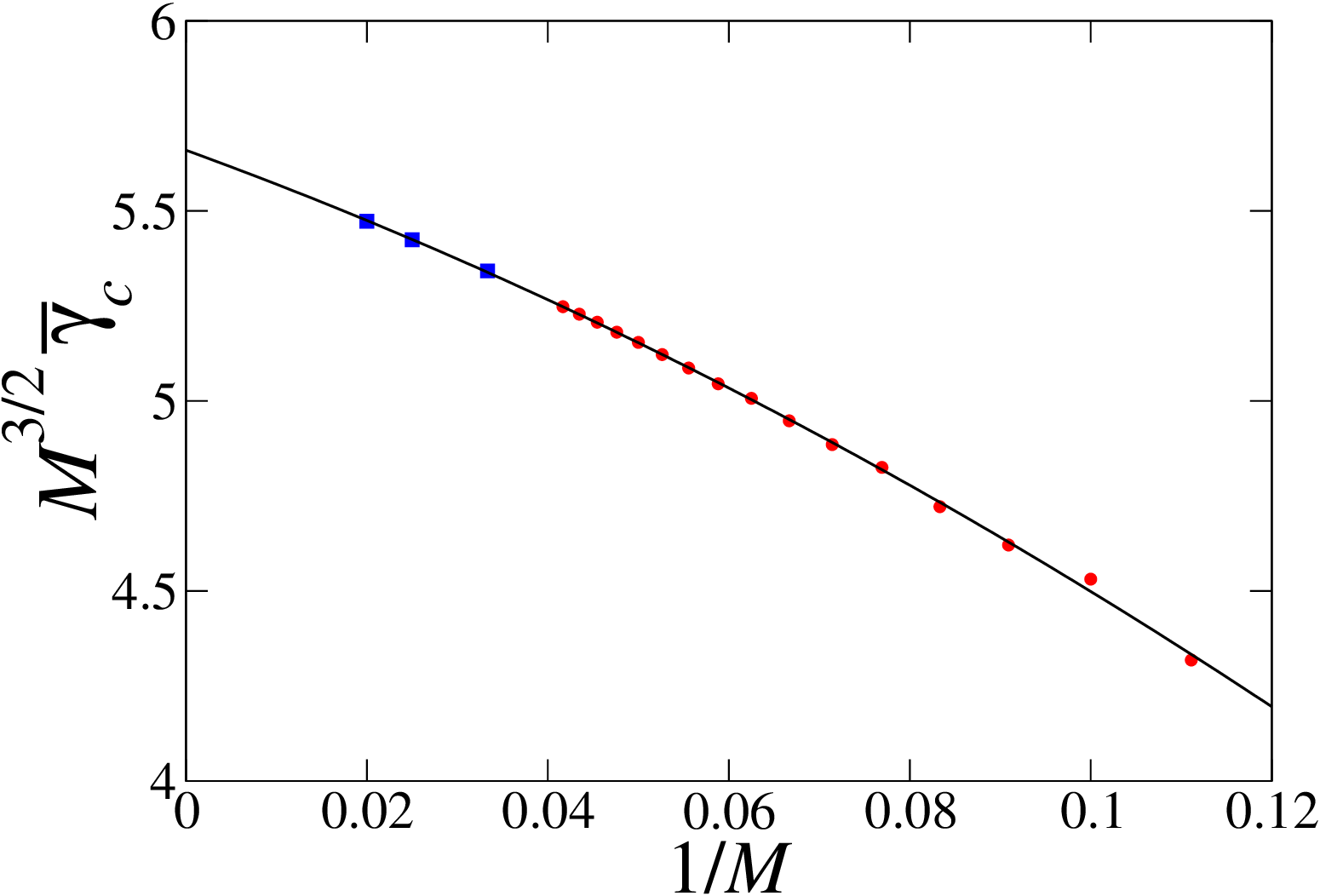}
\caption{\small
Plot of the product $M^{3/2}\bar{\g_c}$ against $1/M$.
Red circles: systematic enumeration of all half-chains up to $M=24$.
Blue squares: numerical simulations with $10^6$ samples.
Black curve: quadratic fit to all data points.}
\label{moyenne}
\end{center}
\end{figure}

Figure~\ref{moyenne} shows the product $M^{3/2}\bar{\g_c}$, plotted against $1/M$.
Red circles show the outcome of a systematic enumeration of all half-chains,
available up to $M=24$.
These data were already presented in Section~\ref{pano}
(see red dataset in Figure~\ref{panorama}).
Blue squares show the outcome of numerical simulations with $10^6$ samples
for $M=30$, 40 and 50.
The black curve is a quadratic fit to all data points with intercept 5.660,
confirming thus the scaling law~(\ref{gmoy}) to a high degree of accuracy,
and yielding the estimate
\beq
A\approx5.7.
\eeq

\begin{figure}[!ht]
\begin{center}
\includegraphics[angle=0,width=.6\linewidth,clip=true]{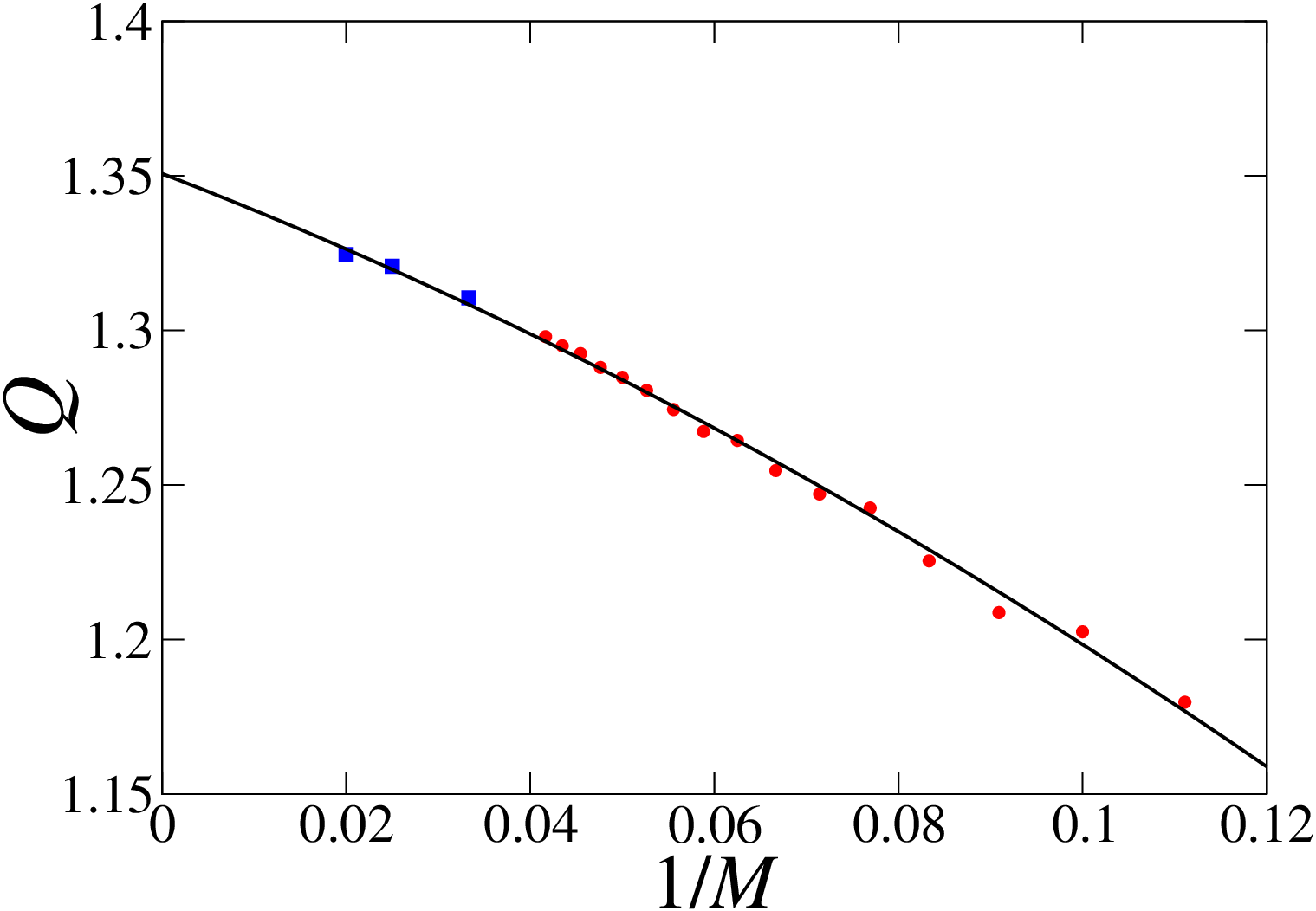}
\caption{\small
Plot of the quantity $Q$ defined in~(\ref{kdef}) against $1/M$.
Same conventions as in Figure~\ref{moyenne}.}
\label{variance}
\end{center}
\end{figure}

Figure~\ref{variance} shows the reduced combination of moments
\beq
Q=\frac{\bar{\g_c^2}}{(\bar{\g_c})^2},
\label{kdef}
\eeq
plotted against $1/M$, with the same conventions as in Figure~\ref{moyenne}.
The black curve is a quadratic fit to all data points with intercept 1.351.
The observed convergence to a well-defined limit
provides a first quantitative piece of information
on the limiting distribution $f(\xi)$,
namely an estimate of its second moment,
\beq
\bar{\xi^2}\approx1.35,\qquad\hbox{i.e.,}\quad\var\xi\approx0.35.
\eeq

\begin{figure}[!ht]
\begin{center}
\includegraphics[angle=0,width=.6\linewidth,clip=true]{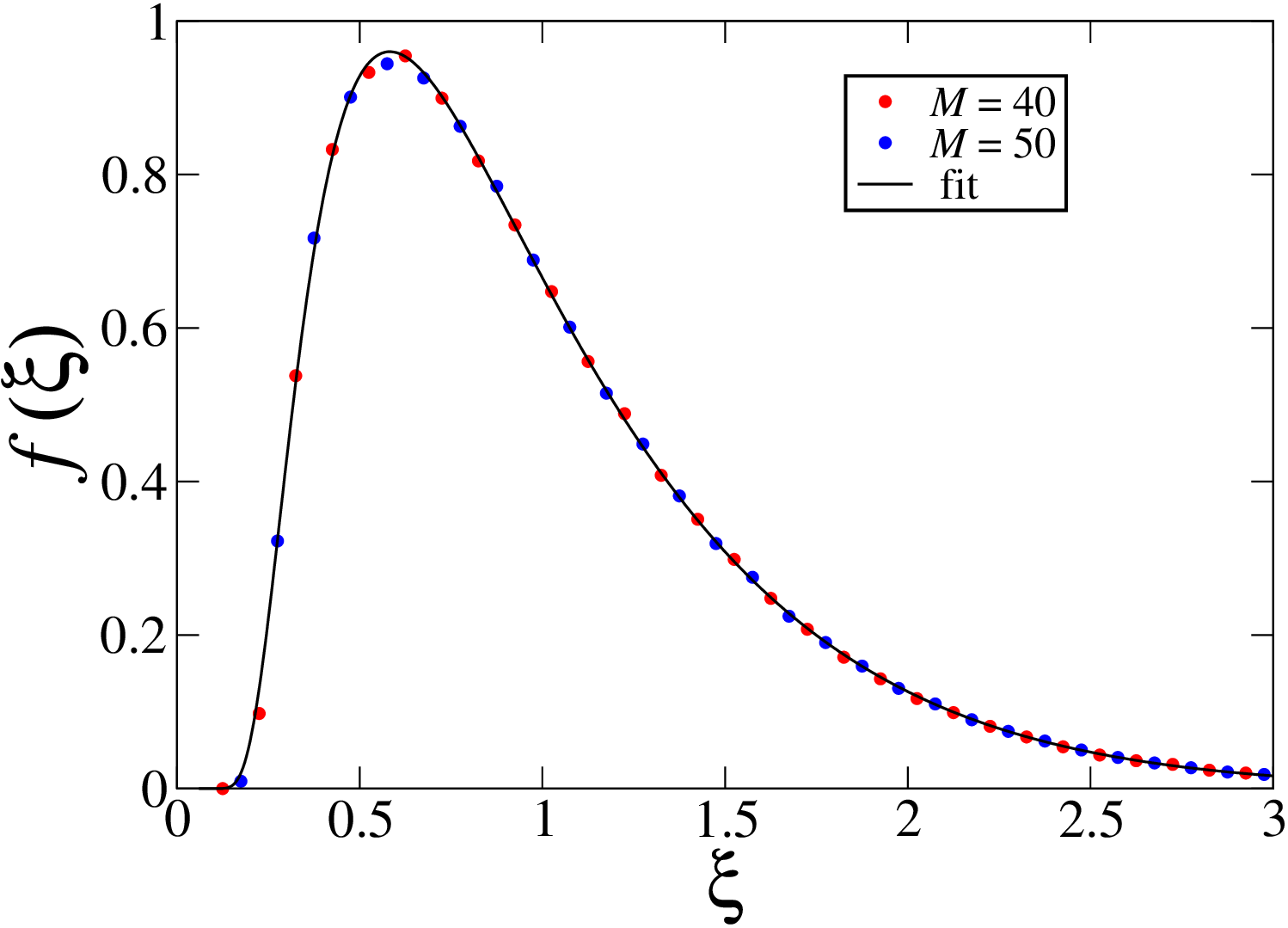}
\caption{\small
Histogram plot of limiting distribution $f(\xi)$ of reduced random variable $\xi$
defined in~(\ref{xidef}),
based on numerical simulations with $10^6$ samples for $M=40$ and 50 (see legend).
Black curve: fit of the form~(\ref{ffit}).}
\label{ghisto}
\end{center}
\end{figure}

Figure~\ref{ghisto} shows a histogram plot
of the distribution of the reduced random variable~$\xi$ defined in~(\ref{xidef}).
Data are based on numerical simulations with $10^6$ samples for $M=40$ and 50
(see legend).
Every second point of each dataset is plotted alternatively.
A good collapse is observed, implying that finite-size corrections are small
for the chosen chain lengths,
and so that the plotted data give an accurate representation of the
limiting distribution $f(\xi)$.
The black curve shows a very accurate four-parameter fit of the form
\beq
f(\xi)\sim\exp\left(-\frac{a}{\xi^2}-b-c\xi-d\xi^2\right).
\label{ffit}
\eeq
This functional form is meant to capture approximately
the asymptotic behavior of the exact distribution at small and large $\xi$.
For a fixed length~$M$,
the smallest threshold is that of diblock sequences (see Section~\ref{diblock}).
We have $\g_c\sim 1/M^2$, and so $\xi\sim1/\sqrt{M}$, with probability $1/2^M$,
suggesting $f(\xi)\sim\exp(-a/\xi^2)$.
The largest threshold, to be investigated in Section~\ref{robust},
scales as $\g_c\sim (\ln M)/M$, and so $\xi\sim\sqrt{M}\,\ln M$,
again with probability $1/2^M$,
suggesting $f(\xi)\sim\exp(-d\xi^2)$, up to logarithmic corrections.
It should however be clear that the exact distribution $f(\xi)$
is expected to be infinitely more complex than~(\ref{ffit}).

\section{Most robust $\PT$-symmetric phase}
\label{robust}

In this section we consider the chains yielding the most robust $\PT$-symmetric phase,
i.e., the highest threshold $\g_c$ at fixed length $M$.
We describe in some detail the outcome
of the systematic investigation described in Section~\ref{pano},
yielding exact data up to $M=24$, i.e., $N=48$.
The value of the threshold,
as well as many other characteristics of these most robust chains,
exhibit an irregular dependence on the length~$M$ of the half-sequence.

\begin{figure}[!ht]
\begin{center}
\includegraphics[angle=0,width=.6\linewidth,clip=true]{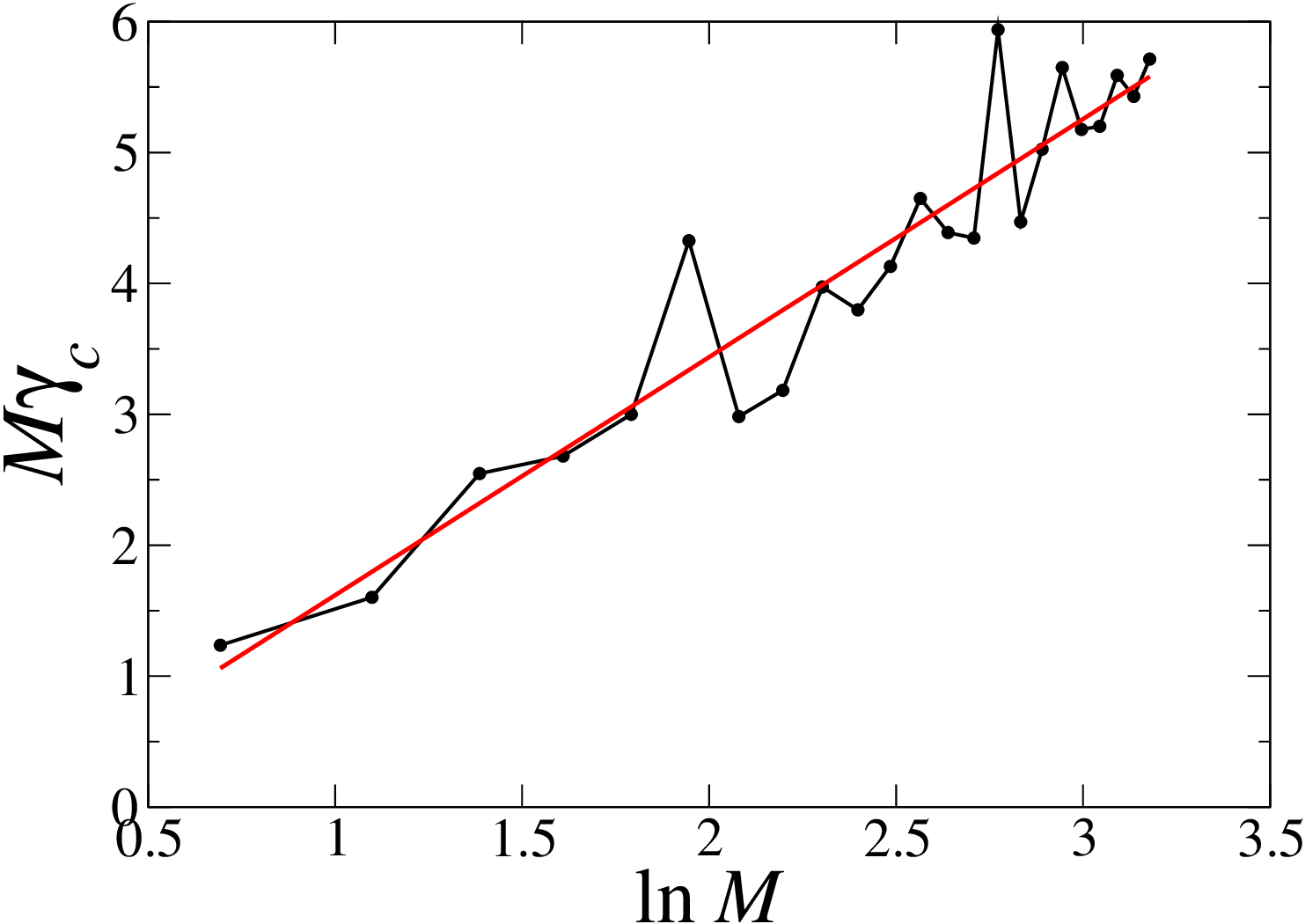}
\caption{\small
Plot of the product $M\g_c$ against $\ln M$,
where $\g_c$ is the highest threshold at fixed length $M$, up to $M=24$.
Red line with slope 1.82: least-square fit.}
\label{robustlog}
\end{center}
\end{figure}

Let us begin with the value of the highest threshold itself.
It has been shown in Section~\ref{alper}
that the thresholds of balanced periodic sequences fall off as $\g_c\approx G_c/M$
(see~(\ref{algal})),
where the amplitude $G_c$ has a complicated dependence
on the length and unit cell of the sequence.
Roughly speaking, the highest threshold $\g_c$ at fixed length~$M$
must be larger than all those thresholds for all periods $L\le M$.
As it turns out, $\g_c$ is larger than $1/M$ by an irregularly varying
but altogether very slowly growing factor.
This is demonstrated in Figure~\ref{robustlog},
showing a plot of $M$ times the highest threshold~$\g_c$ against $\ln M$.
These data were already presented in Section~\ref{pano}
(see blue dataset in Figure~\ref{panorama}).
The red line with slope 1.82 shows an acceptable least-square fit to the data,
suggesting the behavior
\beq
\g_c\approx\frac{C\,\ln M}{M},\qquad C\approx1.8.
\eeq

Table~\ref{sequences} provides a list of all half-sequences
$(\eps_1,\dots,\eps_M)$ yielding the highest threshold shown
in Figure~\ref{robustlog}, for all lengths up to $M=24$,
with the convention $\eps_1=+$ (see Section~\ref{short}).
These sequences can be viewed as consisting of alternating domains $(+-+-\dots)$
separated by domain walls, in the words of magnetism, denoted by vertical bars.
The number of domain walls and their positions optimize the value of the threshold.
We have not succeeded in identifying any pattern in those optimal positions.

In order to illustrate the complexity of the dependence of $\g_c$
on the positions of the domain walls,
we consider in~\ref{appalpha} alternating chains
having a single pair of $\PT$-related domain walls
between sites $K$ and $K+1$ and between sites $2M-K$ and $2M-K+1$.
Already in this situation,
the threshold is found to have an intricate asymptotic dependence
on the ratio $\alpha=K/M$, involving four different regimes.
This threshold is always enhanced by the presence of the domain walls,
with respect to that of pristine alternating chains (see~(\ref{gcalt})).
The enhancement factor takes its maximal value 3 for $\alpha=1/3$.

\begin{table}[!ht]
\begin{center}
\begin{tabular}{|c|c|}
\hline
$M$ & $\eps_1,\dots,\eps_M$\cr
\hline
1 & $+$\cr
2 & $+-$\cr
3 & $+-|-$\cr
4 & $+|+-+$\cr
5 & $+-|-+-$\cr
6 & $\left\{\matrix{+-+|+-|-\cr+-|-+-+\cr+|+-+-+}\right.$\cr
7 & $+-|-+-+-$\cr
8 & $+-|-+-+-+$\cr
9 & $+-|-+-+-+|+$\cr
10 & $+-+|+-+-+-+$\cr
11 & $+-+|+-+-+-+-$\cr
12 & $+|+-+-|-+-+-+-$\cr
13 & $+-+-+-+|+-+-|-+$\cr
14 & $+-+-+-|-+-+|+-|-+$\cr
15 & $+-+|+-|-+-+-+|+-+-$\cr
16 & $\left\{\matrix{+|+-+-+-+-|-|-+|+-+-\cr+|+-+-+-+-|-+|+|+-+-}\right.$\cr
17 & $+-+-+-+|+-|-+|+-+-|-+$\cr
18 & $+-+|+-|-+-+|+-+-+-|-+-$\cr
19 & $\left\{\matrix{+|+-+-+-+-+-|-|-+|+-+-+\cr+|+-+-+-+-+-|-+|+|+-+-+}\right.$\cr
20 & $+|+-+-+-|-+|+-+-|-+|+-+|+-$\cr
21 & $+-|-+|+-|-+|+-+-|-+-+|+-+-+$\cr
22 & $+-+|+-+-+-+-+-|-+|+-+-|-+|+$\cr
23 & $+-+-+-+|+-|-+|+-|-+-+|+-+-|-+$\cr
24 & $+|+-+-+-|-+-+-+-+-+|+-|-+-+|+$\cr
\hline
\end{tabular}
\caption
{Half-sequences yielding the highest thresholds $\g_c$ at fixed length $M$, up to $M=24$.
Vertical bars: domain walls between alternating domains.}
\label{sequences}
\end{center}
\end{table}

Some degeneracies can be seen in Table~\ref{sequences}.
For $M=6$, three different half-sequences yield the maximal $\g_c=1/2$,
with mergings occurring at different critical energies (see below).
For $M=16$ and $M=19$, two different half-sequences yield the same maximal threshold
and mergings at the same critical energies.
The threshold~$\g_c$ generically corresponds to the appearance
of two twofold degenerate eigenvalues at opposite energies $\pm E_c$,
where $E_c>0$ has itself an irregular dependence on $M$ (data not shown).
For $M=2$, 7, 8, 12 and 21,
there is a single twofold degenerate eigenvalue at zero energy.

The case $M=6$ is special in several regards.
Three distinct half-sequences, listed in Table~\ref{sequences},
yield the highest threshold $\g_c=1/2$.
This simple value allows for an analytical study of the corresponding energy spectra.
For the first half-sequence $(+-++--)$, the first mergings occur at $\pm E_c$, with
\beq
E_c=\frac{\sqrt7}{2}\approx1.322\,875.
\eeq
For the second half-sequence $(+--+-+)$, four mergings occur simultaneously
at $\pm E_{c1}$ and $\pm E_{c2}$, with
\beq
E_{c1}=\frac{\sqrt{5-2\sqrt5}}{2}\approx0.363\,271,\quad
E_{c2}=\frac{\sqrt{5+2\sqrt5}}{2}\approx1.538\,841.
\label{e1e2}
\eeq
For the third half-sequence $(++-+-+)$, which is the only unbalanced one,
the first mergings occur at $\pm E_c$, with
\beq
E_c=\frac{\sqrt3}{2}\approx0.866\,025.
\eeq

\section{Discussion}
\label{disc}

This paper has been devoted to non-Hermitian $\PT$-symmetric
tight-binding chains of $N=2M$ sites,
where gain/loss optical potentials of equal magnitudes,
of the form $\ii\g\eps_n$ with $\eps_n=\pm1$,
are distributed over all sites in a periodic, random, or any other fashion.
We have provided a systematic study of the threshold $\g_c$ for $\PT$-symmetry breaking,
with the main emphasis being on the dependence of $\g_c$ on the chain length
for various classes of configurations of optical potentials.
Our main findings are summarized in the
panoramic overview of the asymptotic behavior of $\g_c$ for various classes of chains
given in Section~\ref{pano} (see Figure~\ref{panorama}).
The threshold generically falls off as a power of the chain length, i.e.,
\beq
\g_c\sim\frac{1}{M^\beta}.
\eeq

The decay exponent $\beta$ depends on the configuration of optical potentials.
The largest exponent $\beta=2$ is reached for diblock chains,
and more generally for all unbalanced periodic chains,
i.e., periodic configurations of optical potentials
where each half of the chain has on average either gain or loss.
The smallest exponent $\beta=1$
is attained for alternating chains, and more generally for all balanced periodic chains,
i.e., periodic configurations of optical potentials whose spatial average over one period vanishes.
For random sequences of optical potentials with zero average and finite variance,
the threshold $\gamma_c$ is itself a random variable, whose mean value decays with exponent $\beta=3/2$,
whereas the fluctuations around this mean value obey a universal distribution $f(\xi)$.

These results have been corroborated and made quantitative
by detailed analytical studies performed in Sections~\ref{diper} to~\ref{ran}.
This analysis relies on concepts and techniques from the physics of one-dimensional disordered systems,
including the transfer-matrix approach and the discrete Riccati formalism.

The above results can be put in a broader context as follows.
Consider the sum
\beq
S=\eps_1+\cdots+\eps_M
\eeq
of all reduced optical potentials over the left half of the chain,
and assume that this sum grows as
\beq
S\sim M^\omega,
\label{swander}
\eeq
with a wandering exponent in the range $0\le\omega\le1$.
The three classes of configurations of optical potentials recalled above
respectively yield $\beta=2$, 1 and 3/2,
whereas they clearly obey~(\ref{swander}) with $\omega=1$,~0 and 1/2.
This observation suggests the hyperscaling relation
\beq
\beta=\omega+1.
\label{albe}
\eeq
This relation provides a quantitative form of the rule of thumb
according to which the more the sites with gain and loss are well mixed along the chain,
the more robust the corresponding $\PT$-symmetric phase.
A large class of deterministic aperiodic sequences have non-trivial wandering exponents $\omega$,
besides the three classical values recalled above~\cite{glindex,jmepl,que,bg}.
The relation~(\ref{albe}) certainly holds in those cases as well,
thus predicting a whole zoo of non-trivial asymptotic decay exponents $\beta$.

The chains yielding the most robust ${\cal PT}$-symmetric phase,
i.e., the highest threshold $\gamma_c$ at fixed chain length, have been investigated
by means of a systematic enumeration up to $M=24$, i.e., $N=48$ (see~Section~\ref{robust}).
Both the optimal~$\gamma_c$ itself and the sequences responsible for it
exhibit an irregular dependence on $M$.
The highest threshold presumably falls off asymptotically as $(\ln M)/M$.
These logarithmic corrections somehow represent an exception to the above rule of thumb.
The chains yielding the optimal thresholds indeed contain several domain walls,
i.e., pairs of consecutive sites with gain or loss.
They are therefore slightly less well mixed than pristine alternating ones,
whereas the corresponding thresholds are larger.

To close, it would certainly be interesting
to extend at least some aspects of the present study to more complex situations,
including e.g.~a periodic, quasiperiodic or random modulation
of on-site energies and/or hopping rates,
or quasi-one-dimensional geometries such as ladders,
exhibiting either internal band gaps or flat bands.

\ack
It is a pleasure to thank Philippe Di Francesco for fruitful exchanges.

\appendix

\section{First-order perturbation theory}
\label{apppert}

In this appendix we derive within the framework of first-order perturbation theory
the spectrum of the tight-binding equation
\beq
\p_{n-1}+\p_{n+1}+\ii\g_n\p_n=E\p_n,
\eeq
with Dirichlet boundary conditions,
for an arbitrary sequence of purely imaginary optical potentials $\g_1,\dots,\g_{2M}$.
Our goal is to show that all energy levels remain real within this framework
if and only if the sequence $\g_n$ is $\PT$-symmetric.

In the absence of optical potentials,
the non-degenerate energy eigenvalues and normalized eigenstates read
\beqa
E_a=2\cos\frac{a\pi}{2M+1}\qquad(a=1,\dots,2M),
\label{ea}
\\
\p_{a,n}=\sqrt\frac{2}{2M+1}\sin\frac{an\pi}{2M+1}.
\eeqa
To first order in the optical potentials,
the energy eigenvalues read $E_a+\ii\Gamma_a$, with
\beqa
\Gamma_a&=&\sum_{n=1}^{2M}\g_n\p_{a,n}^2
\nonumber\\
&=&\frac{2}{2M+1}\sum_{n=1}^{M}\left(\g_n+\g_{2M+1-n}\right)\,\sin^2\!\frac{an\pi}{2M+1}.
\label{ga}
\eeqa

If the sequence $\g_1,\dots,\g_{2M}$ is $\PT$-symmetric, obeying
\beq
\g_{2M+1-n}=-\g_n,
\eeq
all the sums in the parentheses in the last line of~(\ref{ga}) vanish,
and therefore all the first-order imaginary parts $\Gamma_a$ vanish,
suggesting the existence of a non-zero threshold~$\g_c$ for $\PT$-symmetry breaking.

The proof of the reciprocal proceeds as follows.
The first-order imaginary parts~(\ref{ga}) obey the mirror symmetry $\Gamma_{2M+1-a}=\Gamma_a$,
so that it is sufficient to consider $a=1,\dots,M$.
Requesting $\Gamma_a=0$ for $a=1,\dots,M$ yields a system of $M$ linear equations
for the $M$ unknown quantities $(\g_n+\g_{2M+1-n})$ for $n=1,\dots,M$.
The corresponding determinant,
\beq
D_M=\det A,
\eeq
with
\beq
A_{an}=\sin^2\!\frac{an\pi}{2M+1}\qquad(a,n=1,\dots,M),
\eeq
is non-zero.
It can indeed be evaluated by using $D_M^2=(\det A)^2=\det(A^2)$.
This is useful since the entries of the matrix $A^2$ have simple expressions~\cite{klm,leq}.
Expanding the sines into complex exponentials and performing the sums,
we indeed obtain after some algebra
\beq
(A^2)_{ab}=\frac{2M+1}{8}\left(1+\half\delta_{ab}\right).
\eeq
The determinant of a matrix of size $M\times M$ with entries $\lambda+\mu\delta_{ab}$
reads $\mu^{M-1}(M\lambda+\mu)$.
We thus obtain
\beq
\abs{D_M}=\frac{(2M+1)^{(M+1)/2}}{2^{2M}}.
\eeq
The sign of $D_M$ is not predicted by this approach.
It depends periodically on $M$~mod~4.

To sum up,
$\Gamma_a=0$ for all $a$ implies $\g_n+\g_{2M+1-n}=0$ for all~$n$,
i.e., $\PT$-symmetry.
This completes the proof.

\section{Chain of length four with arbitrary gains and losses}
\label{appg1g2}

In this appendix we consider the tight-binding equation
\beq
\p_{n-1}+\p_{n+1}+\ii\g_n\p_n=E\p_n
\eeq
on a $\PT$-symmetric chain of length $N=2M=4$,
with arbitrary imaginary optical potentials,
obeying
\beq
\g_3=-\g_2,\qquad\g_4=-\g_1.
\eeq
The first two optical potentials $\g_1$ and $\g_2$ are kept as free parameters.

\begin{figure}[!ht]
\begin{center}
\includegraphics[angle=0,width=.6\linewidth,clip=true]{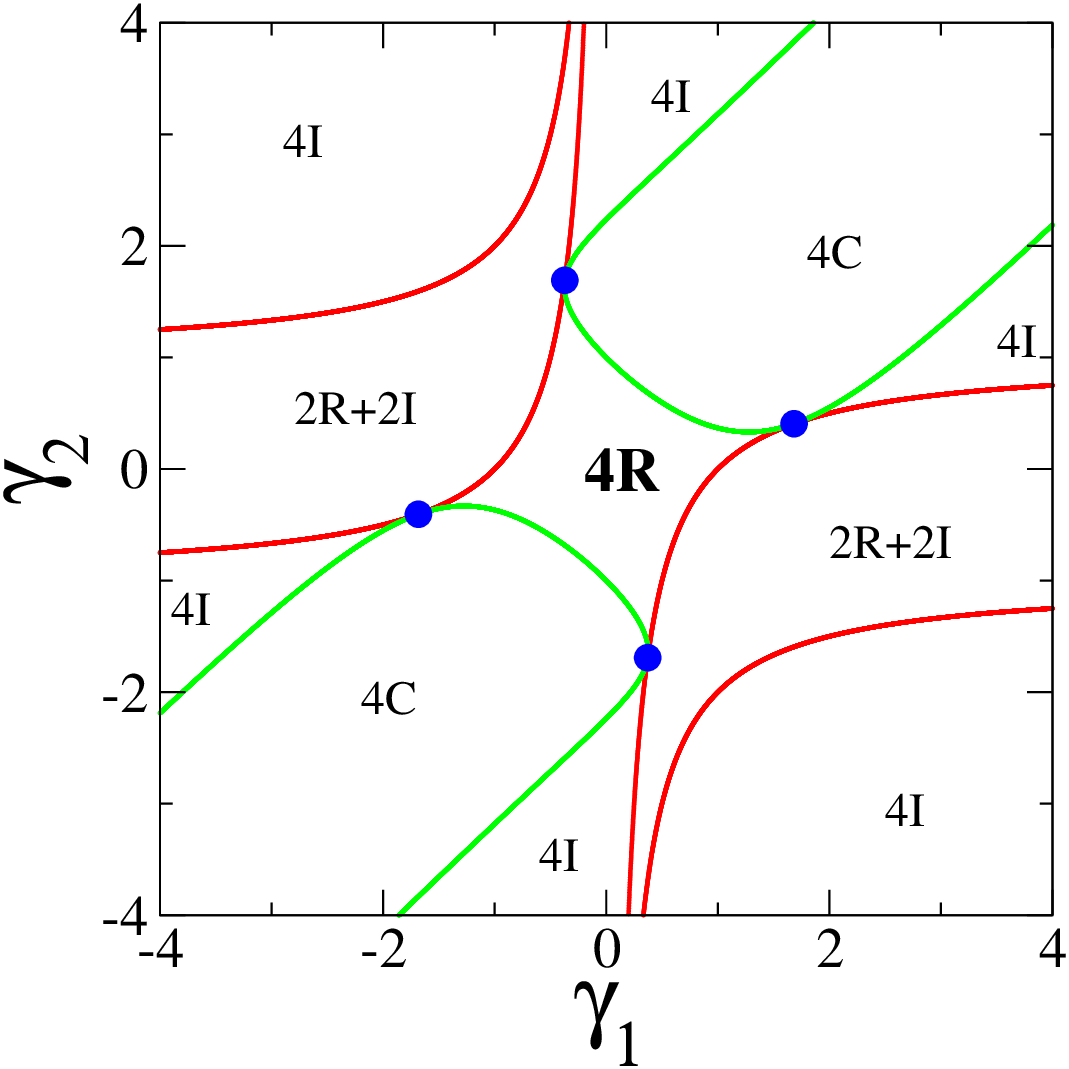}
\caption{\small
Bifurcation diagram of the $\PT$-symmetric chain of length $N=2M=4$
in the $\g_1$--$\g_2$ plane (see text).
$\PT$-symmetry is unbroken in the central region marked 4R.}
\label{g1g2}
\end{center}
\end{figure}

The quantization condition~(\ref{qr}) is expressed by the characteristic polynomial
\beq
P=E^4+(\g_1^2+\g_2^2-3)E^2+(\g_1\g_2+\g_1+1)(\g_1\g_2-\g_1+1).
\eeq
This polynomial has a rich bifurcation diagram shown in Figure~\ref{g1g2}.

There is a twofold degenerate eigenvalue at zero energy on the red curves, i.e., for
\beq
(\g_1\g_2+\g_1+1)(\g_1\g_2-\g_1+1)=0.
\eeq
There are two twofold degenerate eigenvalues at $\pm E_c\ne0$
(either real or imaginary) on the green curves, i.e., for
\beq
(\g_1^2-\g_2^2)^2-2(\g_1+\g_2)(\g_1+3\g_2)+5=0.
\eeq
The red and green curves have four contact points (blue symbols), at
\beqa
(\g_1\approx1.683\,771,\quad\g_2\approx0.406\,095),
\nonumber\\
(\g_1\approx-0.371\,506,\quad\g_2\approx1.691\,739),
\eeqa
and their opposites,
where the characteristic polynomial has a fourfold degenerate eigenvalue at zero energy.
These four points are cocyclic, obeying $\g_1^2+\g_2^2=3$.

The above curves define various regions in the $\g_1$--$\g_2$ plane.
The structure of the energy spectrum is given in each region,
with R, I and C denoting real, imaginary and generic complex eigenvalues.
For instance, 2R+2I means two opposite non-zero real eigenvalues
and two opposite non-zero purely imaginary ones.

$\PT$-symmetry is unbroken whenever all eigenvalues are real,
i.e., in the central region marked 4R.
This quadrangular region is bounded by two red arcs and two green ones
meeting at the four blue contact points.
The predictions~(\ref{gcst2}) and~(\ref{gshort2})
for the thresholds of binary chains with half-sequences $(++)$ and $(+-)$
can be respectively recovered by intersecting one of the green arcs with the bisector line
of equation $\g_2=\g_1$ and of the red ones with the bisector line of equation $\g_2=-\g_1$.

\section{Alternating chains with a single pair of domain walls}
\label{appalpha}

This appendix is devoted to almost pristine alternating chains of length $N=2M$
having a single pair of domain walls at $\PT$-related positions
between sites $K$ and $K+1$ and between sites $2M-K$ and $2M-K+1$.
The corresponding half-sequence reads
\beq
\eps_n=\left\{\matrix{(-1)^n\hfill&(n=1,\dots,K),\hfill\cr
(-1)^{n-1}\hfill&(n=K+1,\dots,M).}\right.
\eeq
For instance, for $M=5$ and $K=2$, the half-sequence reads $(-+|+-+)$,
and so the full sequence reads $(-+|+-+-+-|-+)$,
where vertical bars denote domain walls.

The forthcoming analysis, including notations, closely follows Section~\ref{alt},
which was devoted to pristine alternating chains.
For definiteness we restrict ourselves to the case where $M$ and $K$ are even.
The expressions~(\ref{ralt}) of the Riccati variables still hold for $n=1,\dots,K$.
We have therefore in particular
\beq
R_K=\e^{-\ii\theta}\;\frac{\sin(K+1)q}{\sin Kq}.
\eeq
For $n=K+1,\dots,M$,
the recursion obeyed by the Riccati variables $R_n$
is obtained by changing $\g$ and $\theta$ into their opposites
in~(\ref{rec1}) and~(\ref{rec2}).
In particular, the variables
\beq
Y_n=\frac{R_n-\e^{\ii(\theta-q)}}{R_n-\e^{\ii(\theta+q)}}
\eeq
still obey the recursion~(\ref{recy}).
We have therefore
\beq
Y_M=\e^{2\ii q(M-K)}Y_K.
\eeq
Using the three above equations, we obtain
\beq
R_M=\e^{\ii\theta}\;\frac
{c_M-c_{M+2}+\ii\tan\theta\left(c_M+c_{M+2}-2c_{M-2K}\right)}
{c_{M-1}-c_{M+1}+\ii\tan\theta\left(c_{M-1}+c_{M+1}-2c_{M-2K-1}\right)},
\eeq
with the shorthand notation $c_n=\cos nq$.
Finally, the quantization condition~(\ref{qr}) simplifies to
\beqa
4\sin^2\theta&&\,\sin Kq\,\sin(K+1)q\,\sin(2M-2K)q
\nonumber\\
&&+\sin^2q\,\sin(2M+1)q=0.
\label{qwall}
\eeqa
The above equation is exact for any even integers $M$ and $K$.

\begin{figure}[!ht]
\begin{center}
\includegraphics[angle=0,width=.6\linewidth,clip=true]{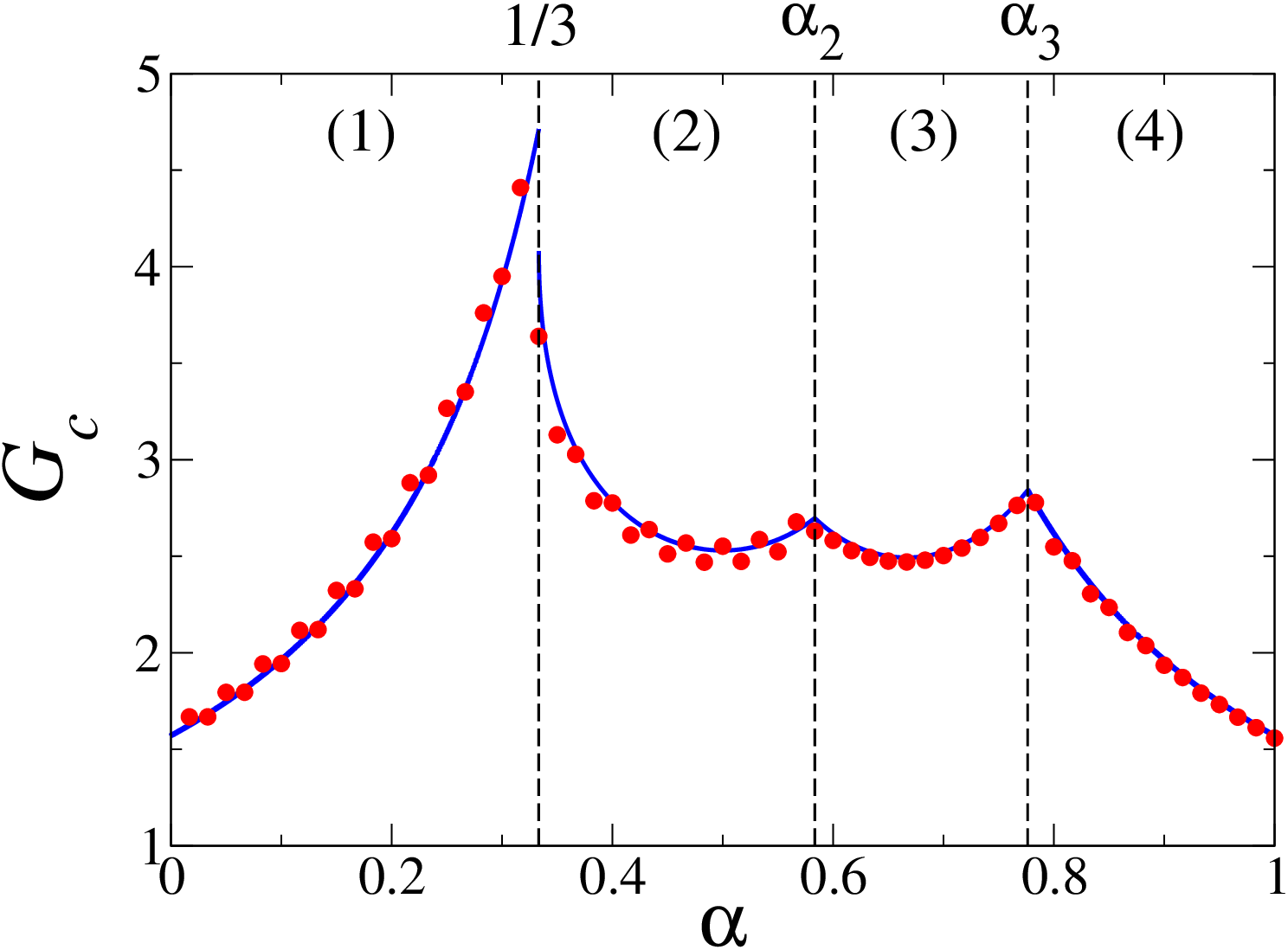}
\caption{\small
Plot of the amplitude $G_c(\alpha)$ against the reduced position $\alpha$ of the left domain wall.
Full curves: analytical asymptotic predictions (\ref{gca1}), (\ref{gca2}), (\ref{gca3}), (\ref{gca4})
in the four regimes.
Symbols: data for $M=60$ and all possible positions of the domain walls.}
\label{galpha}
\end{center}
\end{figure}

For long chains, where both $M$ and $K$ are large,
the threshold for $\PT$-symmetry breaking scales as
\beq
\g_c\approx\frac{G_c(\alpha)}{M},
\eeq
where the numerator has a rich dependence on the ratio
\beq
\alpha=\frac{K}{M},
\eeq
as announced in Section~\ref{robust}.
The amplitude $G_c(\alpha)$ is always larger
than that of pristine alternating chains (see~(\ref{gcalt})),
which is recovered for $\alpha\to0$ and $\alpha\to1$.
The effect of the domain walls indeed disappears in both limits,
where they either attain the endpoints of the chain
or annihilate each other in the middle.
The threshold amplification is maximal for $\alpha\to(1/3)_-$,
where we have $G_c((1/3)_-)/G_c(0)=3$.
Figure~\ref{galpha} shows a plot of $G_c(\alpha)$ against~$\alpha$.
Numerical results for the thresholds
for $M=60$ and all possible positions of the domain walls
are compared to the exact expression of $G_c(\alpha)$,
involving four different regimes, to be successively investigated below.

\begin{itemize}

\item
Regime~(1) holds for $0<\alpha<1/3$.
The amplitude increases from $G_c(0)=\pi/2\approx1.570\,796$
to $G_c((1/3)_-)=3\pi/2\approx4.712\,388$.

\item
Regime~(2) holds for $1/3<\alpha<\alpha_2$.
At $\alpha=1/3$ the amplitude drops discontinuously
to $G_c((1/3)_+)=3\pi\sqrt{3}/4\approx4.081\,048$.
We have $\alpha_2\approx0.583\,417$ and $G_c(\alpha_2)\approx2.697\,312$.

\item
Regime~(3) holds for $\alpha_2<\alpha<\alpha_3$.
The amplitude is continuous at $\alpha_2$ and at $\alpha_3$,
with $\alpha_3\approx0.776\,365$ and $G_c(\alpha_3)\approx2.841\,880$.

\item
Regime~(4) holds for $\alpha_3<\alpha<1$.
The amplitude decreases from $G_c(\alpha_3)$ to $G_c(1)=\pi/2\approx1.570\,796$.

\end{itemize}

Analytical expressions of the amplitude $G_c(\alpha)$ in each regime
can be derived by zooming onto the exact quantization condition~(\ref{qwall}) in appropriate ranges.

In Regimes~(1) and~(4), the first merging takes place at zero energy.
This corresponds to $\theta=\pi/2$,
whereas $q=\pi/2-\eta$, and so $\gamma_c\approx2\eta$, with $\eta\sim1/M$.
The quantization condition~(\ref{qwall}) asymptotically becomes
\beq
\cos(1-2\alpha)G_c=0.
\eeq
The amplitude $G_c(\alpha)$ is the smallest positive solution to the above equation.
We thus obtain the explicit expressions
\beqa
G_c(\alpha)=\frac{\pi}{2(1-2\alpha)}\quad\hbox{in Regime~(1)},
\label{gca1}
\\
G_c(\alpha)=\frac{\pi}{2(2\alpha-1)}\quad\hbox{in Regime~(4)}.
\label{gca4}
\eeqa

In Regime~(2), the first mergings take place near zero energy.
This corresponds to $q=\pi/2-\eta$ with $\eta\sim1/M$ and $\theta$ variable,
and so $E_c\approx2\eta\cos\theta$ and $\g_c\approx2\eta\sin\theta$.
Introducing the product $x=2M\eta$,
the quantization condition~(\ref{qwall}) asymptotically becomes
\beq
2\sin^2\theta\,\sin\alpha x\,\sin(1-\alpha)x+\cos x=0.
\eeq
The threshold is obtained by the maximization procedure applied to~(\ref{gxcurve}), yielding
\beq
G_c(\alpha)=\stackunder{\max}{x}
x\sqrt{-\frac{\cos x}{2\sin\alpha x\,\sin(1-\alpha)x}},
\label{gca2}
\eeq
where the maximum is taken over $x$ in the range $\pi/2<x<3\pi/2$ at fixed~$\alpha$.
The amplitude takes its minimal value $G_c(1/2)\approx2.529\,882$ for $\alpha=1/2$, i.e.,
when the left domain wall is in the middle of the half-sequence.
It has a square-root endpoint singularity at $\alpha=1/3$, below which Regime~(1) takes over.
Setting
\beq
\alpha=\frac13+\delta,
\eeq
the maximum takes place for
\beq
x\approx\frac{3\pi}{2}-\frac{3\pi\sqrt{3\delta}}{4}
\eeq
and reads
\beq
G_c\approx\frac{3\pi\sqrt{3}}{4}-\frac{9\pi\sqrt\delta}{4}.
\eeq

In Regime~(3), the first mergings take place near the band edges.
This corresponds to both $q$ and $\theta$ scaling as $1/M$.
Introducing the product $x=Mq$,
we obtain the asymptotic prediction
\beq
G_c(\alpha)=\stackunder{\max}{x}
\frac{x}{\sin\alpha x}\sqrt{-\frac{\sin 2x}{\sin 2(1-\alpha)x}},
\label{gca3}
\eeq
where the maximum is taken over $x$ in the range $\pi/2<x<\pi$ at fixed~$\alpha$.
The amplitude takes its minimal value $G_c(2/3)\approx2.493\,125$ for $\alpha=2/3$.

\section*{References}

\bibliography{paper.bib}

\end{document}